\documentclass[5p,final]{elsarticle}

\biboptions{numbers,sort&compress}

\usepackage[T1]{fontenc}
\usepackage{libertine}
\usepackage{libertinust1math}

\usepackage{amsmath}
\usepackage{bbold}
\usepackage{graphicx}
\usepackage{eurosym}
\usepackage{mathtools}
\usepackage{url}
\usepackage{booktabs}
\usepackage{epstopdf}
\usepackage{xfrac}
\usepackage{tabularx}
\usepackage{bm}
\usepackage{subcaption}
\usepackage{blindtext}
\usepackage{longtable}
\usepackage{multirow}
\usepackage{threeparttable}
\usepackage{pdflscape}
\usepackage[export]{adjustbox}
\usepackage[version=4]{mhchem}
\usepackage[colorlinks]{hyperref}
\usepackage[parfill]{parskip}
\usepackage[nameinlink]{cleveref}
\usepackage[leftcaption,raggedright]{sidecap}
\usepackage[prependcaption,textsize=footnotesize]{todonotes}
\usepackage{enumitem} 
\usepackage{todonotes}
\usepackage{tabularx}
\usepackage{xurl}

\makeatletter
\renewcommand \thesection{S\@arabic\c@section}
\renewcommand\thetable{S\@arabic\c@table}
\renewcommand \thefigure{S\@arabic\c@figure}
\def\ps@pprintTitle{%
  \let\@oddhead\@empty
  \let\@evenhead\@empty
  \let\@oddfoot\@empty
  \let\@evenfoot\@empty
}
\def\ps@pprintTitle{%
  \let\@oddhead\@empty
  \let\@evenhead\@empty
  \let\@oddfoot\@empty
  \let\@evenfoot\@empty
}

\makeatother

\usepackage{siunitx}
\DeclareSIUnit\year{a}
\DeclareSIUnit{\tco}{t_{\ce{CO2}}}
\DeclareSIUnit{\sieuro}{\mbox{\euro}}

\usepackage{lipsum}

\usepackage[resetlabels,labeled]{multibib}
\newcites{S}{Supplementary References}
\bibliographystyleS{elsarticle-num}

\usepackage[nolist]{acronym}

\newcommand\rev[1]{#1}

\graphicspath{
    {../Optimal Design of Large Thermal Energy Storage/figures/},
}

\newcommand{\N}{\mathcal{N}}
\newcommand{\T}{\mathcal{T}}

\newcommand{\timesteps}{\mathcal{T}}

\newcommand{\charge}{q^{\text{charge}}}
\newcommand{\discharge}{q^{\text{discharge}}}

\newcommand{\maxSoc}{\overline{\soc}}
\newcommand{\maxVolume}{v^{\text{max}}}
\newcommand{\soc}{e}
\newcommand{\standingLosses}{\eta^{\text{stand}}}

\newcommand{\eToP}{\mu^{\text{e2p}}}
\newcommand{\heatCapacity}{c_p}
\newcommand{\topTemp}{T^{\text{top}}}
\newcommand{\bottomTemp}{T^{\text{bottom}}}
\newcommand{\retTemp}{T^{\text{ret}}}
\newcommand{\fwdTemp}{T^{\text{fwd}}}

\newcommand{\volumeFlow}{\dot{V}}
\newcommand{\density}{\rho}
\newcommand{\alphaRH}{\alpha^\text{RH}}
\newcommand{\alphaRet}{\alpha^\text{ret}}
\newcommand{\alphaHpThIn}{\alpha^\text{HP-src}}
\newcommand{\Qsource}{\dot{Q}^\text{discharge}}
\newcommand{\Qboost}{\dot{Q}^\text{RH}}
\newcommand{\Qret}{\dot{Q}^\text{HP-src}}
\newcommand{\powerRH}{q^\text{RH}}

\newcommand{\thInputHP}{q^\text{HP-src}}

\begin{document}

\begin{acronym}
    \acro{COP}{coefficient of performance}
    \acroplural{COP}[COPs]{coefficients of performance}
	\acro{LDES}{long duration energy storage}
    \acro{LTES}{large thermal energy storage}
    \acro{TES}{thermal energy storage}
    \acro{PTES}{pit thermal energy storage}
	\acro{ATES}{aquifer thermal energy storage}
    \acro{TTES}{tank thermal energy storage}
    \acro{BTES}{borehole thermal energy storage}
    \acro{ASHP}{air-sourced heat pump}
    \acroplural{ASHP}[ASHPs]{air-sourced heat pumps}
    \acro{GSHP}{geothermal-sourced heat pump}
    \acroplural{GSHP}[GSHPs]{geothermal-sourced heat pumps}
	\acro{WSHP}{water-sourced heat pump}
	\acroplural{WSHP}[WSHPs]{water-sourced heat pumps}
	\acro{RWSHP}{river water-sourced heat pump}
	\acroplural{RWSHP}[RWSHPs]{river water-sourced heat pumps}
	\acro{SWSHP}{sea water-sourced heat pump}
	\acroplural{SWSHP}[SWSHPs]{sea water-sourced heat pumps}
	\acro{EHSHP}{excess heat-sourced heat pump}
	\acroplural{EHSHP}[EHSHPs]{excess heat-sourced heat pumps}
    \acro{CHP}{combined heat-and-power}
    \acro{VRES}{variable renewable energy sources}
    \acro{LAU}{local administrative unit}
    \acro{OCGT}{open-cycle gas turbine}
    \acroplural{OCGT}[OCGTs]{open-cycle gas turbines}
    \acro{CCGT}{closed-cycle gas turbine}
    \acroplural{CCGT}[CCGTs]{closed-cycle gas turbines}
    \acro{CAPEX}{capital expenditures}
	\acro{OPEX}{operational expenditures}
	\acro{CC}{carbon capture}
	\acro{DAC}{direct air capture}
	\acro{BECCS}{bioenergy with carbon capture and storage}
	\acro{SOC}{state of charge}
\end{acronym}

\newcommand{\noPTES}{\emph{No PTES~}}
\newcommand{\freeBoosting}{\emph{Free Boosting~}}
\newcommand{\freeCapacity}{\emph{Free Capacity~}}
\newcommand{\resistiveBoosting}{\emph{Resistive Boosting~}}
\newcommand{\hpBoosting}{\emph{Heat Pump Boosting~}}

\newcommand{\lowTemp}{\emph{LowST~}}
\newcommand{\midTemp}{\emph{MidST~}}
\newcommand{\highTemp}{\emph{HighST~}}

\newcommand{\HP}{\emph{heat pump}}
\newcommand{\HPs}{\emph{heat pumps}}
\newcommand{\ASHP}{\emph{air-sourced \HP}}
\newcommand{\ASHPs}{\emph{air-sourced \HPs}}
\newcommand{\RWSHP}{\emph{river-water-sourced \HP}}
\newcommand{\RWSHPs}{\emph{river-water-sourced \HPs}}
\newcommand{\SWSHP}{\emph{sea-water-sourced \HP}}
\newcommand{\SWSHPs}{\emph{sea-water-sourced \HPs}}
\newcommand{\GSHP}{\emph{geothermal-sourced \HP}}
\newcommand{\GSHPs}{\emph{geothermal-sourced \HPs}}
\newcommand{\EHSHP}{\emph{excess-heat-sourced \HP}}
\newcommand{\EHSHPs}{\emph{excess-heat-sourced \HPs}}

\newcommand{\red}[1]{\textcolor{red}{[#1]}}

\begin{frontmatter}

	\title{Mind the Temperature Gap: The Role of Pit Thermal Energy Storage in a Sector-Coupled Energy System with High-Temperature District Heating}
    
	\author[affil]{Caspar Schauß} \corref{correspondingauthor}
	\ead{c.schauss@tu-berlin.de}
	\author[affil]{Amos Schledorn}
	\ead{a.schledorn@tu-berlin.de}
	\author[affil]{Tom Kähler}
	\ead{t.kaehler@tu-berlin.de}
	\author[hi]{Kristina Schumacher}
	\ead{schumacher@hamburg-institut.com}
	\author[hi]{Mathias Ammon}
	\ead{ammon@hamburg-institut.com}
	\author[affil]{Tom Brown}
	\ead{t.brown@tu-berlin.de}

	\cortext[correspondingauthor]{Corresponding author}

	\address[affil]{ Department of Digital Transformation in Energy Systems, Institute of Energy Technology, Technische Universität Berlin, Fakultät III, Einsteinufer 25 (TA 8), 10587 Berlin, Germany }
	\address[hi]{HIR Hamburg Institut Research gGmbH, Paul-Nevermann-Platz 5, 22765 Hamburg, Germany}

	\begin{abstract}
		Pit thermal energy storage (PTES) provides large-scale thermal storage capacity in district heating systems, supporting flexibility on both daily and seasonal scales. Most existing large-scale energy system studies on PTES do not account for temperature differences between storage and the network. Neglecting these temperature differences can result in less efficient PTES integration, since they affect usable energy capacity and introduce additional costs for discharge requiring temperature boosting.

To explore how temperature constraints shape the system-level value of PTES we use PyPSA-DE, an open-source sector-coupled capacity expansion model of Germany and neighbouring countries in a scenario with net-zero carbon emissions for 2045. The model explicitly represents Germany's 40 largest district heating systems with multiple heat sources—including geothermal, river and sea water as well as electrolyzer waste heat pumps. To isolate PTES effects, we examine counterfactual scenarios: systems without PTES, idealized systems with PTES but without temperature constraints, and feasible systems with resistive or heat-pump boosting.

We find that PTES reduces German annual system costs by 135--345~M€~a$^{-1}$ relative to systems relying solely on tank storage. Lowering maximum forward temperatures from 124~°C to 95~°C decreases district-heating costs by 7.6\% without PTES and 10\% with PTES. Idealized scenarios without temperature constraints yield district heating cost savings of up to 15\%, indicating that temperature-agnostic modelling overestimates PTES benefits. PTES provides economic value even under current high temperatures, though temperature misalignment limits its contribution during peak demand due to the need for boosting. The findings highlight PTES' role in leveraging low-price electricity through electrified heating while emphasizing the importance of explicitly accounting for temperature constraints.
	\end{abstract}


\end{frontmatter}

\section{Introduction}
\label{sec:intro}

As Germany and Europe pursue climate neutrality by mid-century, district heating systems are increasingly recognised as key infrastructures for reducing emissions and integrating renewable energy in urban heat supply on a European \cite{DirectiveEU20232023} and national level \cite{Germany_HeatPlanningAct_2023, Germany_BuildingsEnergyAct_2023, bmwk-bundesministerium_fur_wirtschaft_und_klimaschutz_mehr_2023}. In district heating systems, \ac{TES} is a key enabler of flexibility, allowing heat supply and demand to be decoupled and improving sector integration with the electricity system \cite{christensen_role_2024, sifnaios_impact_2023, brown_synergies_2018, gea-bermudezRoleSectorCoupling2021, bogdanovLowcostRenewableElectricity2021}.

\ac{TES} technologies in district heating systems differ in scale, storage medium, and site requirements. \Ac{TTES} uses steel or concrete vessels and typically provides short-term storage (hours to days) \cite{sifnaios_impact_2023} for volumes of 10$^3$–5*10$^4$ m$^3$ \cite{ieaestask39LargeThermalEnergy2024}. \Ac{PTES} consists of large, water-filled excavations with liners and insulated lids, scaling up to 10$^5$–2*10$^6$ m$^3$ for inter-weekly to seasonal operation \cite{sifnaios_impact_2023}. \Ac{ATES} and \ac{BTES} employ subsurface formations as storage media and can reach volumetric capacities of up to 10$^6$ and 5*10$^6$ respectively \cite{ieaestask39LargeThermalEnergy2024}, but both involve complex subsurface requirements and extensive pre-investigation efforts \cite{fleuchausWorldwideApplicationAquifer2018}.
Among them, \ac{PTES} has attracted particular attention because of its comparatively low specific capital costs at large volumes and high charge/discharge power ratings \cite{danishenergyagencyTechnologyDataEnergy2025} that enable both seasonal shifting \cite{sifnaios_dronninglund_2023} and daily balancing \cite{sifnaiosPerformanceAnalysisHoje2025}.

These favorable attributes suggest that \ac{PTES} could deliver substantial savings for low-emission energy systems facilitating \ac{VRES} integration \cite{sifnaios_impact_2023}, reducing peak electricity needs for electrified heating applications \cite{zeyenMitigatingHeatDemand2021}, and serve as a complementary, even competing long-duration energy storage alternative to hydrogen storage to provide seasonal flexbility \cite{victoria_role_2019, schmidtMixLongdurationHydrogen2025}.

\begin{figure}
    \centering
    \includegraphics[width=0.4\textwidth]{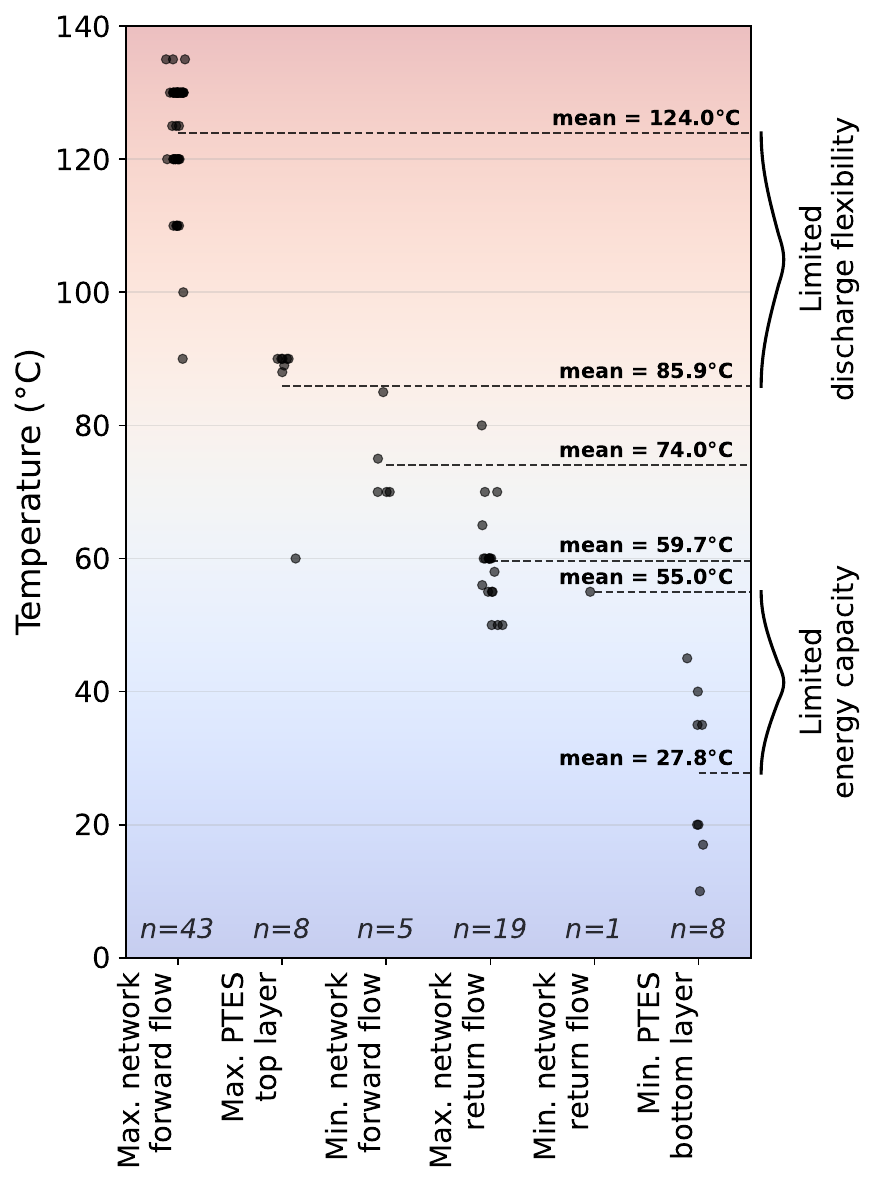}
    \caption{Common network temperatures in largest German district heating systems \cite{pelda_district_2021} and typical \ac{PTES} temperature ranges \cite{danishenergyagencyTechnologyDataEnergy2025} at implemented Danish sites. The temperature gaps indicate the need for temperature boosting when discharging \ac{PTES} into legacy networks and considering network return temperatures for an accurate estimation of usable energy capacity.}
    \label{fig:ptes_network_interaction}
\end{figure}

Realising these benefits, however, depends on the compatibility between the network and storage temperatures that govern how efficiently heat can be charged, stored, and discharged. In \ac{PTES}, thermal stratification leads to distinct hot and cold layers, with temperatures typically ranging from $90\,^{\circ}$C at the top to $10$–$35\,^{\circ}$C at the bottom \cite{danishenergyagencyTechnologyDataEnergy2025}. Many European district heating systems built to first- to third-generation standards, by contrast, still operate with forward flow temperatures up to $130\,^{\circ}$C \cite{euroheat_2024} and return flow temperatures above $50\,^{\circ}$C \cite{bertelsenLoweringTempsExisting2025,pelda_district_2021}.
Although lowering network temperatures enables a more efficient integration of renewable heat sources \cite{manz_spatial_2024} and heat pumps \cite{viethDistrictHeatingNetwork2025} as well as a reduction of thermal losses \cite{schmidt_low_temperature_2017}, the high associated cost of retrofitting existing infrastructure \cite{averfalkLowTemperatureDistrictHeating2021} and modernizing the building stock \cite{euroheat_2024} remains a major barrier to implementation.
At such conditions, the interaction between \ac{PTES} and network temperatures are shaped by two fundamental effects, which are illustrated in \cref{fig:ptes_network_interaction}.

First, discharging into networks with supply temperatures exceeding the \ac{PTES} top temperature requires temperature boosting, i.e., auxiliary energy supply to lift \ac{PTES} outlet temperatures to the required network supply levels \cite{sorknaes_simulation_2018}. 
Although boosting technically enables discharge to higher-temperature sinks, the associated auxiliary energy demand—particularly as the \ac{PTES}-network temperature gap widens in winter when \ac{VRES} output is scarce \cite{pieper_assessment_2019}—introduces additional costs.
These can reduce the effective discharge flexibility from an economic perspective relative to \ac{TES} that operate at or above district heating supply temperature levels.

Second, the available energy capacity $\maxSoc$ of \ac{PTES} depends directly on the achievable temperature spread within the storage. According to \cref{eq:energy_capacity} \cite{sorknaes_simulation_2018}, it can be expressed as the product of the storage volume $\maxVolume$, the specific heat capacity $\heatCapacity$, the density of water $\density$, and the temperature difference between top and bottom layers:
\begin{equation}
\maxSoc = (\topTemp - \bottomTemp) \cdot \heatCapacity \cdot \density \cdot \maxVolume
\label{eq:energy_capacity}
\end{equation}
The network’s return temperature constrains the storage bottom temperature and thereby the exploitable temperature spread. A smaller spread reduces the energy capacity per cubic meter and increases the specific \ac{CAPEX} per kWh \cite{danishenergyagencyTechnologyDataEnergy2025}. Boosting \ac{PTES} with a heat pump not only enables discharge into high-temperature networks but can also extend the usable temperature spread by simultaneously chilling the bottom temperature, thereby increasing the maximum energy capacity of the storage at the same volumetric capacity \cite{sorknaes_simulation_2018}.

A broad body of literature investigates the integration of \ac{TES} in large-scale energy systems and the implications of varying district heating network temperatures. System-level analyses consistently highlight the value of \ac{TES} in reducing system costs and enhancing flexibility in future energy systems. Zeyen et al.~\cite{zeyenMitigatingHeatDemand2021} demonstrate that \ac{TES} can mitigate peak demands, reducing total system costs by up to 3\% and limiting the need for gas boiler capacity in a 2050 European net-zero context. Similarly, Brown et al.~\cite{brown_synergies_2018} find that including district heating thermal storage in an interconnected European sector-coupled energy system decreases market prices by 14\% and total system costs by 7\%. Both studies use the European capacity expansion model PyPSA-Eur, which is comparable to the model introduced later in this paper, but apply a simplified district heating formulation. \ac{TES} is represented with cost assumptions below 35~€/m$^{3}$, indicative of \ac{PTES}, and heat pump \acp{COP} are computed using a fixed supply temperature of 55~°C without explicit modeling of return temperatures. Consequently, tem\-perature-dependent aspects of \ac{TES} integration in district heating networks are not addressed.

Kök et al. \cite{kokAchievingClimateNeutrality2025} apply the Enertile model in a greefield setting to assess temperature reduction scenarios but restrict analysis to supply temperatures below 95~°C. The study shows that lower network temperatures can reduce levelized costs of heat generation by 20\% and promote a heat-pump-dominated mix. Billerbeck et al.\cite{billerbeck_integrating_2024}, using the same model, find that lower network temperatures increase renewable heat potential and efficiency while reducing biomass dependency; however, even the high-temperature scenario assumes supply temperatures of only around 80~°C. Neither study includes \ac{PTES} nor examines higher-temperature networks typical of 1G–3G systems.
Moreover, several other sector-coupled energy system models consider district heating with water tanks \cite{abdelkhalekPyPSAEarthSectorcoupledGlobal2025, sanchezdieguezModellingDecarbonisationTransition2021} or generic \ac{TES} \cite{hantoAssessingImplicationsHydrogen2024, maierImpactForesightHorizons2025} but lack explicit \ac{PTES} representation.

A few system-level studies have included \ac{PTES} explicitly. Sifnaios et al.~\cite{sifnaios_impact_2023} employ the Balmorel model to analyze the role of \ac{PTES} and \ac{TTES} in a Denmark-centered, sector-coupled energy system. They find that including \ac{PTES} can lower district heating prices by up to 10\,\% and that only scenarios with \ac{TES} reach climate neutrality, as \ac{PTES} enables more efficient integration of variable renewables. However, their temperature assumptions ($\Delta T = 90$–40\,°C) reflect Danish conditions, not higher-temperature systems. While they perform a \ac{CAPEX} sensitivity suggesting reduced \ac{PTES} benefits at higher operating temperatures, other tem\-per\-a\-ture-re\-la\-ted impacts—such as increased boosting requirements and reduced heat-pump performance—are not captured.  
Gea-Bermudez et al.~\cite{gea-bermudezRoleSectorCoupling2021} also apply the Balmorel model to analyze the role of sector coupling in a net-zero Northern Central European system. They find that \ac{PTES} provides seasonal flexibility, reduces investments in district heating generation capacity, and facilitates the integration of variable renewables. However, tem\-per\-a\-ture-de\-pen\-dent integration constraints of \ac{PTES} are not represented.

More detailed small-scale modeling approaches have focused on individual district heating systems to capture \ac{PTES} dynamics more accurately. Sorknæs et al.~\cite{sorknaes_simulation_2018} present a detailed simulation model of \ac{PTES} co-operated with an absorption booster heat pump, which serves as a reference for capturing operational interactions not representable in system-scale models. Sporleder et al.~\cite{SPORLEDER2024100564} link a mixed-integer linear optimization model with detailed simulations to optimize \ac{PTES} investments in a German district heating system under varying temperature scenarios, explicitly accounting for boosting requirements and tem\-per\-a\-ture-de\-pen\-dent storage capacities. Sifnaios et al.~\cite{SIFNAIOS2025127770} use a TRNSYS simulation to evaluate \ac{PTES} operation in a Danish network, finding that it reduces the cost of heat by 14\,\%. Geyer et al.~\cite{geyerEnergyeconomicAssessmentReduced2021} quantify the economic impact of reducing network temperatures for various storage technologies and report cost reduction gradients of 2.33\,€/(MWh\,°C) for \ac{PTES}, further highlighting the importance of temperature-aware modeling at the district heating level.

Against this background, a gap remains between large-scale energy system studies--which, reflecting modelling trade-offs between spatial aggregation and operational detail \cite{kotzurModelersGuideHandle2021}, typically assume lower district heating network temperatures and represent thermal storage in a simplified form--and detailed system models, which capture tem-per-a-ture-dependent storage behaviour but do not assess cross-sectoral and cross-regional interactions. In particular, the implications of \ac{PTES} in national-scale energy systems with higher-temperature networks, where supply temperatures exceed 90~°C and discharge temperatures together with boosting requirements affect both storage efficiency and electricity demand, are not yet well understood.

Accordingly, this work addresses the following research questions that, for the first time in large-scale energy system modeling, explicitly account for temperature-dependent \ac{PTES} integration constraints:
\begin{itemize}
\item What is the role of \ac{PTES} in large-scale sector-coupled energy systems?
\item How is the role of \ac{PTES} affected by more accurate modelling of temperature in district heating?
\item How is this role affected by varying district heating temperatures and \ac{PTES} boosting regimes in a net-zero German energy system?
\end{itemize}

To answer these questions, this work makes three distinct contributions:
\begin{itemize}
\item it extends the district heating representation of a large-scale, sector-coupled energy system model with high spatial resolution by explicitly introducing forward and return network temperatures;
\item it links these temperatures to the operation of heat pumps and \ac{PTES}, accounting for temperature-dependent energy capacity and boosting requirements; and
\item it provides a quantitative analysis of the system-level role of \ac{PTES} under realistic, yet previously unexplored, district heating temperature regimes in a net-zero German energy system.
\end{itemize}

The remainder of the paper is structured as follows. 
\Cref{sec:methods} introduces PyPSA-DE and describes the representation of future district heating systems, heat pumps, and \ac{PTES}, including tem\-per\-a\-ture-de\-pen\-dent boosting constraints. It also describes the investigated scenarios. 
\Cref{sec:results} reports results, followed by a discussion and conclusions in \cref{sec:discussion,sec:conclusion}, respectively.

\section{Methods}
\label{sec:methods}

We take PyPSA-DE\cite{pypsa-de}, an existing open-source energy system model of Germany embedding in Europe and introduce several improvements to its representation of district heating including \ac{PTES} and heat source modelling as well as explicitly resolving the 40 largest German district heating systems.

PyPSA-DE\footnote{\href{https://github.com/PyPSA/pypsa-de}{https://github.com/PyPSA/pypsa-de}} builds upon the European model PyPSA-Eur\footnote{\href{https://github.com/PyPSA/pypsa-eur}{https://github.com/PyPSA/pypsa-eur}}. This section focuses on the representation of district heating in PyPSA-DE. More in-depth descriptions of PyPSA-DE and PyPSA-Eur can be found in \cite{pypsa-de} as well as the supplementary materials of \cite{neumann_potential_2023} and \cite{victoria_speed_2022}.

\subsection{PyPSA-DE}
PyPSA-DE solves a linear optimization problem that minimizes annualized investment and operation costs. It enforces an energy balance that ensures demand is met at all times, locations, and for all carriers; the dual variable of this constraint is interpreted as the price of the respective carrier in time and space. The model is further subject to technical and physical constraints, such as a carbon budget and hourly \ac{VRES} availability. The latter is derived from 2019 weather data \cite{c3s_era5_2018} using \textit{atlite} \cite{hofmann_atlite_2021}. PyPSA-DE is configured to achieve carbon neutrality by 2045.

In addition to Germany, it covers Austria, Belgium, Switzerland, Czech Republic, Denmark, France, UK, Luxembourg, Netherlands, Norway, Poland, Sweden, Spain, and Italy to consider cross-border energy flows.
The model comprises 49 regions (30 for Germany, 3 for Italy, 2 each for Denmark, the UK and Spain, 1 each for Austria, Belgium, Switzerland, Czech Republic, France, Luxembourg, the Netherlands, Poland and Sweden and Norway) at a 3-hourly temporal resolution, which has been shown to be a good compromise between accuracy and runtime \cite{neumann_potential_2023}.
It covers the demand sectors households, services, industry, and transport, and includes a broad set of energy carriers such as electricity, heat, hydrogen, methane, oil, biomass, and methanol. 
The material flow of carbon dioxide is additionally represented to account for emissions and carbon management, including capture, storage, and sequestration in processes like fossil fuel use, fuel synthesis, \ac{DAC}, industrial \ac{CC}, and \ac{BECCS}.
In this study, we apply a brownfield approach towards 2045 carbon neutrality, which retains those of today's electricity generation and transmission capacities, including \ac{CHP} plants, with lifetimes exceeding 2045.

\subsection{District heating in PyPSA-DE}
\subsubsection{Spatial resolution}
We model 30 German base regions for simulating the power sector and networks for electricity, hydrogen, and CO$_2$. The 40 largest German district heating systems, according to demand from the German district heating atlas \cite{pelda_district_2021}, are then modelled as district heating sub-regions of the respective base regions in which they are located.
District heating demand and supply in the base regions represents all district heating outside the 40 largest networks.
To retain tractable model sizes, the sub-regions share the base region's connections for the generation and transmission of electricity, natural gas, biomass and hydrogen, including excess heat from electrolysis. Only district heating is supplied and consumed separately per base and sub-region. The modelling of individual heating remains unaltered.

Land eligibility for \ac{PTES} and heat sources is restricted to the respective base/sub-region. 
We assume the sub-regions to be geographically restricted to the district heating area of the respective city, which was derived from public census data of 2022 \cite{destatis_zensus2022_heizungsart_wohnungen}. 
For the base regions, district heating areas for land eligibility are based on future projected district heating areas published in \cite{manz_spatial_2024}.

Heat can be transported freely within each base and sub-region but not across base/sub-regions. Thus, network topology is only considered implicitly through grid losses as a constant percentage increase in demand. For district heating sub-regions, grid losses are estimated based on the annual feed-in data German district heating atlas \cite{pelda_district_2021}. Grid losses in other systems are assumed to be 15\% of the heat demand in each time step, average losses in Germany \cite{pelda_district_2021}. 

\subsubsection{District heat demand}
The total national final energy demand for heating applications is derived from JRC-IDEES \cite{rozsai_jrc-idees-2021_2024} and Eurostat \cite{eurostat_database_nodate} data, as well as additional statistics for Switzerland \cite{bfe_swiss_federal_office_for_energy_energieverbrauch_nodate} and Norway \cite{statistics_norway_energy_2014}. National district heating shares are also derived from these statistics.
District heating can be used to cover heat demand of the residential and services sector as well as low-temperature industrial heat.
Within each country, heat demand in the residential and services sector is distributed to model regions based on population density assuming a district heating market share of 40\% in urban areas. 
For all countries except Germany, district heating heat demand is estimated based on the national district heating share and the ratio of urban population per model region.
Low-temperature industrial heat demand is distributed according to the locations of industrial sites.
For the 40 district heating sub-regions, demand is assigned according to the German district heating atlas \cite{pelda_district_2021} and subtracted from the containing model region accordingly.

Daily heat load factors are derived based on daily average ambient temperatures \cite{c3s_era5_2018} in the respective model region and heating degree day approximation assuming a threshold temperature of 15\textdegree C \cite{hofmann_atlite_2021}.
\rev{Hourly demand profiles from the German Association of Energy and Water Industries (BDEW) \cite{oemof_developer_group_oemofdemandlib_2025}}, specific to the service and residential sectors as well as to weekdays and weekends, are used to compute hourly heat demand. Low-temperature industrial heat demand is assumed constant.

\subsubsection{District heat supply}
PyPSA-DE allows district heating demand to be supplied through several technologies. Power-to-heat units include resistive heaters and heat pumps using different heat sources, namely air, river water, sea water, geothermal heat, and excess heat from electrolysis plants.
Investment costs of some district heating technologies are listed in \cref{tab:capex}. Most other technology cost are based on \cite{danishenergyagencyTechnologyDataEnergy2025, danishenergyagencyTechnologyDataGeneration2025}.

Geothermal heat potentials at 65 °C \cite{manz_spatial_2024} are mapped to district heating areas in each (sub-region) assuming a buffer of 5km.
For deriving the river heat potentials, data on river discharge and ambient temperatures from HERA \cite{tilloy_hera_2024} are used. For the mapping of the ambient temperature to the river water temperature a regression by \cite{triebs_fernwaerme_2023} is applied. 
We further assume a minimum outlet temperature of 1 °C and a maximum cooling of 1 K of the river's entire volume flow at a recovery distance of 25 km \cite{triebs_fernwaerme_2023,jung_estimation_2020} to compute the thermal energy of rivers.
The data on sea surface temperature by the Copernicus Marine Service \cite{copernicus_cmems_moi_00170_med_subskin_sst} were used to estimate the sea water heat potentials in the coastal regions and sub-regions. Sea water is assumed to be an inexhaustible heat source, although a minimum inlet temperature of 1 °C is assumed.
To map river-water and sea-water heat potentials to (sub-)regions, we apply the same buffer of 5km as buffer as for geothermal heat sources.

Furthermore, \ac{CHP} generation fired by natural gas, hydrogen, waste and biomass as well as natural-gas and biomass-fired boilers are available. Today's \ac{CHP} capacities are assigned to district heating systems using the German Marktstammdatenregister, an official register for plants participating in the electricity and gas market, maintained by the Federal Network Agency \cite{hulk_open-mastr_2023}. 
Biogenic resources across all sectors are limited to residual products from agriculture and forestry based on the medium scenario of the JRC ENSPRESO database \cite{ruiz_enspreso_2019}.
Oil and gas resources are restricted to the capacities at European extraction sites.
We further enforce waste incineration equal to a heat output of 18.5~TWh across Germany.

\subsubsection{Large thermal energy storage}
Heat can be stored in \ac{TTES} and \ac{PTES}, whereas \ac{PTES} investments are admitted in Germany only to isolate the effect of introducing \ac{PTES} in Germany.
There are no limits to the the maximum installable capacity for \ac{TTES} due to its significantly smaller footprint \cite{danishenergyagencyTechnologyDataEnergy2025}.

\ac{PTES}, on the other hand, requires significant land, often sparse in dense urban areas \cite{danishenergyagencyTechnologyDataEnergy2025}. Thus, its expansion is limited by the total eligible area per model region. Eligible areas must be categorized as arable land, pastures, shrub/herbaceous, or open spaces with little vegetation according to the OpenStreetMap landcover data \cite{OSMLC}. Furthermore, they must be outside protected areas according to the NATURA 2000 guidelines \cite{european_environment_agency_natura_2024} and have a groundwater depth $\ge 10$\,m \cite{fan_hydrologic_2017}. The eligible area is translated to storage potential in m$^3$ assuming an area requirement of 25,000 m$^2$ for a \ac{PTES} with a volume of 70,000 m$^3$, which aligns with common storage parameters \cite{xiang_comprehensive_2022,danishenergyagencyTechnologyDataEnergy2025}. Therefore, among the areas computed through intersecting the data on land cover, protected areas and groundwater depth, only the ones larger than 25,000 m$^2$ are kept.

\subsubsection{Temperature modelling}\label{sec:methods-temperatures}
Ambient temperatures, network temperatures and heat pump \acp{COP} are resolved in time and space. 
We assume a piecewise linear relationship between ambient and forward temperature \cite{pieper_assessment_2019}. The forward temperature remains at its minimum value for ambient temperatures above 10\textdegree C. Below 10\textdegree C, it increases linearly with decreasing ambient temperatures and reaches its maximum value for ambient temperatures lower or equal than 0\textdegree C.

For all countries but Germany, forward temperatures from the Euroheat DHC Market Outlook \cite{euroheat_2024} are used, with maximum annual forward temperatures reaching from 75\textdegree C in Denmark to 130\textdegree C in Czech Republic and Poland today. 
In Germany, we assume a maximum forward temperature of 124\textdegree C based on data reported by district heating companies \cite{pelda_district_2021} and return flow temperatures of 60\textdegree C \cite{agfw_hauptbericht}. 
Countries with missing data are assumed to operate at the average network temperatures of the reported countries.
Network temperature differences across district heating systems in the same country are not accounted for.

For this study, where we focus on \ac{PTES} in Germany, we limit forward temperatures to values greater-or-equal than 90\,\si{\celsius}, which we assume to be the constant temperature of the \ac{PTES} top layer (see \cref{sec:ptes-model}).
This ignores higher heat pump \acp{COP} and direct heat-source utilisation in the summer but allows to model \ac{PTES} operation more realistically. If we were to include lower forward temperatures, our assumption of constant \ac{PTES} top temperatures would be less realistic, as the model would be allowed to store heat generated at temperatures below 90\textdegree C and discharge \ac{PTES} at 90\textdegree C in winter undermining boosting requirements, which would be unphysical.

Heat pump \acp{COP} for district heating are approximated thermodynamically based on the temperature delta between heat source (air, river, sea, geothermal or electrolyser excess heat) and sink (network forward) temperatures \cite{jensen_heat_2018} at a source cooling of 6 K. 
This is the heat source cooling for the \ac{COP} computation only, thus differing from the admitted cooling of the entire volume flow of rivers of 1 K assumed in their heat source computation.
We assume ammonia as a heat pump refrigerant and no heat losses but an isentropic compressor efficiency of 80\% \cite{pieper_comparison_2020}.

\subsection{PTES model} \label{sec:ptes-model}
This section introduces those parts of the formulation of PyPSA-DE which are unique to the present study. 
The mathematical formulation of PyPSA-DE itself is largely equivalent to that of PyPSA-Eur, for a formulation of which the reader is referred to \cite{brown_synergies_2018}.

We account for the dependence of the energetic storage capacity of \ac{PTES} on the temperature of the stored water as well as for boosting requirements to overcome the temperature difference between the stored water and forward flow. 
Temperature levels in the \ac{PTES}, as well as forward and return temperatures (\cref{sec:methods-temperatures}), are set exogenously to maintain a linear optimisation problem.

We assume a constant fixed top temperature $\topTemp$ of 90\textdegree C. In reality, the top temperature in \ac{PTES} systems varies and depends on storage operation, measurements from existing systems \cite{sifnaios_monitoring_2025} indicate that sustaining $\approx\!$ 90\textdegree C is feasible under frequent charging.
By this, we inherently assume short-term rather than seasonal storage operation at which constant temperatures might not be realistic.

Temperature boosting of the \ac{PTES} discharge flow is enabled either via resistive heaters or heat pumps. In general, system integrations via other higher-temperature heat generators, such as boilers and \ac{CHP}, would also feasible but outside the scope of this study.
When boosting via resistive heaters, the assumed constant \ac{PTES} bottom temperature $\bottomTemp$ is set to the temperature of the network's return flow $\retTemp$.
When boosting via heat pumps is enabled, heat pump operation is assumed to achieve bottom temperatures of 10\textdegree C by utilising that part of the \ac{PTES} discharge heat flow at temperatures below the return temperature as a heat source. 
Data from current \ac{PTES} projects indicate that the bottom temperature is typically around 35\textdegree C \cite{danishenergyagencyTechnologyDataEnergy2025} (see \cref{fig:ptes_network_interaction}), but lower values may occur after extended periods of heat pump operation \cite{sifnaios_dronninglund_2023}. 
A schematic of \ac{PTES} integration is illustrated in \cref{fig:ptes_schematic}.

\subsubsection{Energetic storage capacity}\label{sec:energetic-capacity}
\ac{PTES} investment are determined through the product of investment cost per m$^3$ and storage volume $\maxVolume_n$ in [m$^3$] in (sub-)region $n \in \N$. In other words, varying the storage temperature delta scales investment costs per energetic capacity, as 
the energetic storage capacity $\maxSoc_n$ is given by the product of the volumetric heat capacity $\heatCapacity$ [kWh/m$^3$/K], the storage volume $\maxVolume_n$, and the temperature delta $(\topTemp_n - \bottomTemp_n)$ between the top and bottom layer temperatures (\cref{eq:ptes-max-soc}). 
\Cref{eq:ptes-soc-limits} sets the corresponding bounds for the energetic \ac{SOC} $\soc_{n,t}$.
Its evolution is governed by \cref{eq:ptes-balance} through standing losses $\standingLosses$, charge ($\charge$) and discharge ($\discharge$) power.
The energy-to-power ratio $\eToP$ sets the permitted peak charge and discharge rates in relation to the volumetric storage capacity (\cref{eq:e2p-ratio}).
\begin{subequations}\label{eq:ptes_soc}
\begin{align}
    & \maxSoc_n =  (\topTemp_n - \bottomTemp_n)\,\heatCapacity\,\maxVolume_n \quad & \forall n \in \N, t \in \timesteps \label{eq:ptes-max-soc}\\
    & 0 \le \soc_{n,t} \le \maxSoc_n \quad & \forall n \in \N t \in \timesteps \label{eq:ptes-soc-limits}\\
    & \soc_{n,t} = (1-\standingLosses) \, \soc_{n,t-1} \nonumber\\ & + \charge_{n,t} - \discharge_{n,t} & \forall n \in \N t \in \timesteps  \setminus \{1\}\label{eq:ptes-balance} \\
    & \charge_{n,t}, \discharge_{n,t} \leq \eToP \, \maxVolume_n & \forall n \in \N, t \in \timesteps \label{eq:e2p-ratio}\\
    & \soc_{n,t}, \charge_{n,t}, \discharge_{n,t} \in \mathbb{R}^+ & \forall n \in \N t \in \mathcal{T} \\
    & \maxVolume_n, \maxSoc_n \in \mathbb{R}^+ & \forall n \in \N
\end{align}
\end{subequations}

While the standing losses depend on geometry, storage temperatures, operation pattern, cycle duration and weather conditions, we assume constant relative losses of 0.18\%/day \cite{danishenergyagencyTechnologyDataEnergy2025}.

\subsubsection{Boosting via resistive heaters}\label{sec:resistive-boosting}

\begin{figure*}[ht]
  \centering
  \begin{subfigure}[t]{0.49\textwidth}
    \centering
  \includegraphics[width=\linewidth,height=0.30\textheight,keepaspectratio]{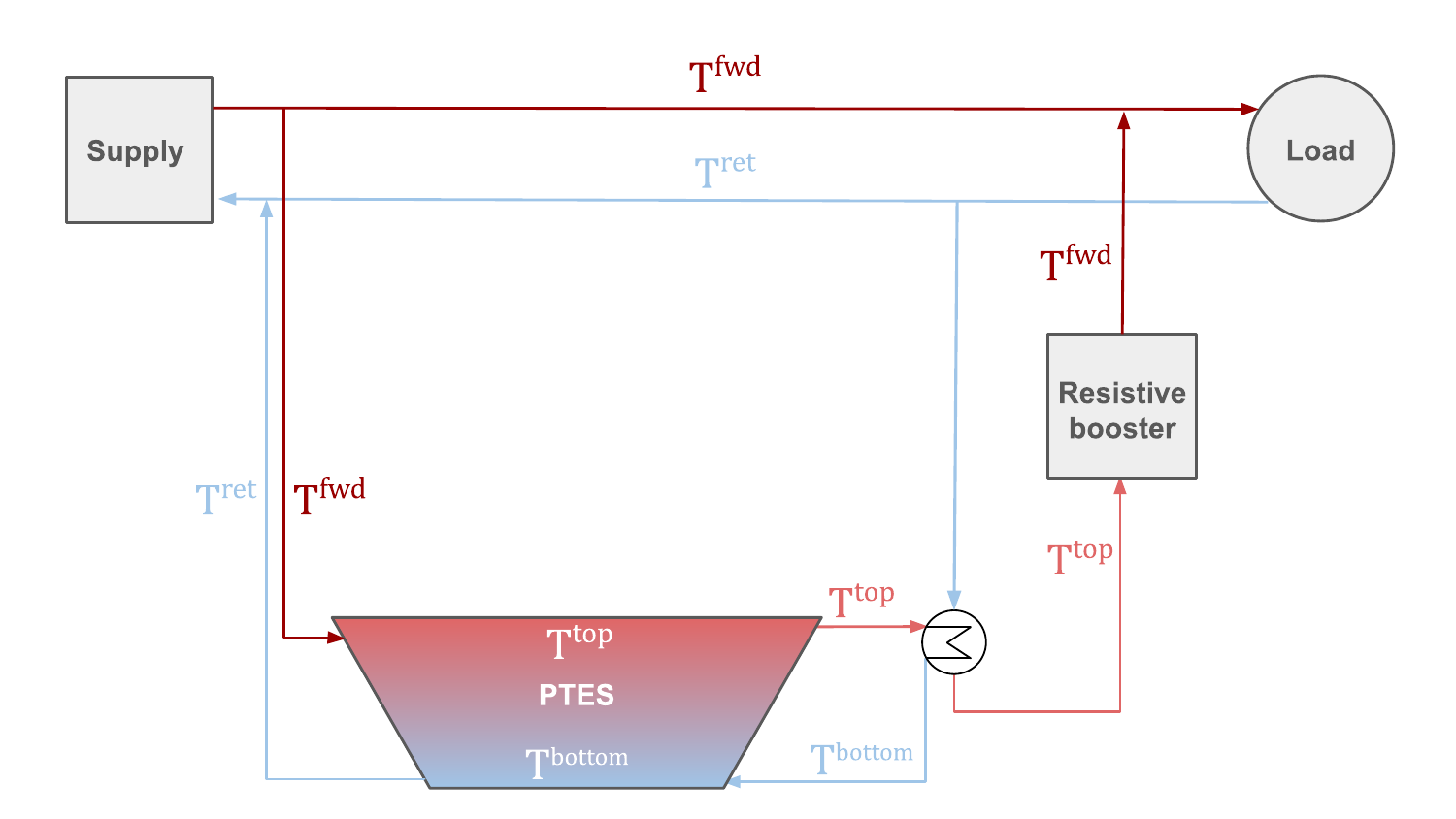}
    \caption{PTES boosted by resistive heater}
  \end{subfigure}\hfill
  \begin{subfigure}[t]{0.49\textwidth}
    \centering
  \includegraphics[width=\linewidth,height=0.30\textheight,keepaspectratio]{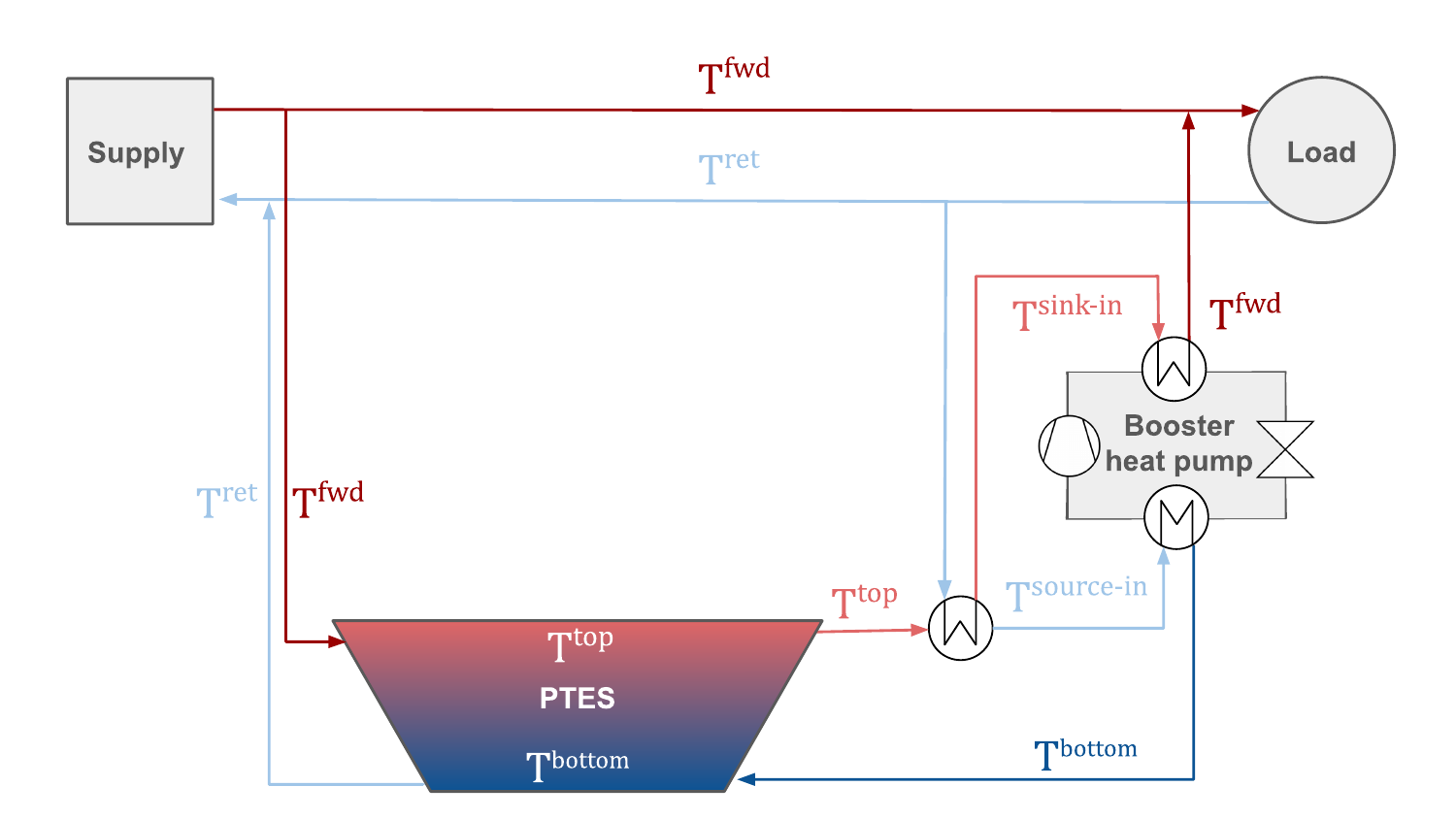}
    \caption{PTES boosted by heat pump}
  \end{subfigure}
  \caption{Schematic integration of \ac{PTES} into district heating systems with (a) resistive booster and (b) booster heat pump. Potential mass flows (arrows) and corresponding temperatures (color-coded) are shown for a state where $\fwdTemp_t > \topTemp$.}
  \label{fig:ptes_schematic}
\end{figure*}

In continuous time\footnote{We assume, here and elsewhere, perfect heat exchangers.}, the \ac{PTES} output heat flow $\Qsource_t$ is the product of the volumetric flow $\volumeFlow_t$, density $\density$, heat capacity $\heatCapacity$ and the spread between storage top layer and resulting bottom temperature (\cref{eq:source-power}).
Note that, when boosting via resistive heaters, we assume $\bottomTemp = \retTemp$.
Additional boosting, denoted as $\Qboost_t$ is required if the forward temperature exceeds the top layer temperature. 
Assuming the volume flow constant across charging and boosting, this can be written as \cref{eq:boost-power}\footnote{\label{fn:indicator_function}$\mathbb{1}(x)$ is the indicator function, such that $\mathbb{1}(x) = \begin{cases}1, & \text{if } x \text{ true}\\0, & \text{otherwise}\end{cases}$}.
\begin{subequations}
\begin{align}
    & \Qsource_t = \volumeFlow_t \, \density \, \heatCapacity \, (\topTemp - \bottomTemp) \quad \label{eq:source-power}\\
    & \Qboost_t =
    \volumeFlow_t \, \density \, \heatCapacity \, (\fwdTemp_t - \topTemp) \, \mathbb{1}(\topTemp < \fwdTemp_t) \label{eq:boost-power}
\end{align}
\end{subequations}

For the optimization model, in discrete time, we compute a-priori the required boost energy per unit of \ac{PTES} discharge  $\alphaRH_{n,t}$ as the ratio of the additional boost and storage output in each (sub\-)region $n \in \N$ and time step $t \in \T$ (\cref{eq:boosting-ratio-rh}). 
We then enforce in the optimization model the heat generation of resistive heaters $\powerRH_{n,t}$ to be at least $\alphaRH_{n,t}$ units of heat for every unit of storage discharge (\cref{eq:boosting-constraint-rh}).
\begin{subequations}
\begin{align}
& \alphaRH_{n,t} = 
    \begin{cases}
      \dfrac{\fwdTemp_{n,t} - \topTemp_n}{\topTemp_n - \bottomTemp_n}, & \topTemp_n < \fwdTemp_{n,t}\\
      0, & \topTemp_{n,t} \geq \fwdTemp_{n,t}
    \end{cases} &\forall n \in \N, t \in \T\label{eq:boosting-ratio-rh}\\
  & \powerRH_{n,t}
  \;\ge\;
  \alphaRH_{n,t} \, \discharge_{n,t} \, 
  \quad & \quad \forall n \in \N, t \in \T \label{eq:boosting-constraint-rh}\\
  & \powerRH_{n,t}, \discharge_{n,t} \in \mathbb{R}^+ & \forall n \in \N, t \in \T
\end{align}
\end{subequations}
This formulation allows the resistive heater to deliver heat beyond the boosting demand, thereby enabling it to operate as both a booster and as a stand-alone supply unit.

\subsubsection{Boosting via heat pumps}\label{sec:hp-boosting}
Booster heat pumps use the storage outflow, pre-cooled to the network return flow, as a heat source, which is assumed to be at a constant temperature $\retTemp$ (\cref{fig:ptes_schematic}):
The \ac{PTES} system would be operated such that during discharge, the storage outflow is used to heat the network return flow from $\retTemp$ to $\topTemp$ and the storage output flow is cooled from $\topTemp$ to $\retTemp$, assuming ideal heat exchangers.
The pre-cooled storage outlow at $\retTemp$ is the heat pump source inlet, which further cools that stream to $\bottomTemp=10\text{\textdegree C}$.

As for other heat sources, the \ac{COP} is calculated based on \cite{jensen_heat_2018}. Here, we set sink inlet and outlet temperatures to $\topTemp$ and $\fwdTemp_t$ and source inlet and outlet temperatures to $\retTemp$ and $\retTemp-6\text{K}$ respectively.
The source outlet temperature assumed for the \ac{COP} approximation is above $\bottomTemp=10\text{\textdegree C}$, which is used in \cref{eq:ptes-max-soc,eq:preheat-power}.
This is because in reality, \ac{PTES} would repeatedly serve as a heat source for several heat pump cycles with a combined cooling beyond 6K
\footnote{Still, assuming constant PTES top and bottom temperatures is a simplification. In reality, the top temperature could fall below 90 \textdegree C over time and bottom temperatures could be higher or lower than 10 \textdegree C, depending on operation and losses. To maintain a linear optimisation problem, we neglect these hydraulic details.}. 
Hence, booster heat pumps reduce $\bottomTemp$ below $\retTemp$ and increase storage energy density (energetic capacity per volumetric capacity), effectively reducing energy-specific CAPEX (see \cref{eq:ptes-max-soc}).

Similar to the case of boosting via resistive heaters, the output heat flow of the storage unit itself $\Qsource_t$ is set by the difference between top and bottom temperature (\cref{eq:ptes-power}) in continuous time. 
However, since, $\bottomTemp=10\text{\textdegree C} < \retTemp$, that temperature delta is larger. 
We calculate heat pump boosting through its source inlet heat flow $\Qret_t$ and the temperature difference between source inlet $\retTemp$ and outlet $\bottomTemp$ (\cref{eq:preheat-power}).
\begin{subequations}
\begin{align}
  & \Qsource_t = \volumeFlow_t \, \density \, \heatCapacity \, (\topTemp - \bottomTemp)  \label{eq:ptes-power}\\
  & \Qret_t = \volumeFlow_t \, \density \, \heatCapacity \, (\retTemp - \bottomTemp) \, \mathbb{1}(\topTemp < \fwdTemp_t) \label{eq:preheat-power}
\end{align}
\end{subequations}

For the PyPSA model in discrete time, the ratio of these heat flows $\alphaHpThIn_{n,t}$ is computed a-priori as a parameter.
Then, the heat pump source inlet flow $\thInputHP_{n,t}$ is set accordingly (\cref{eq:boosting-ratio-hp}). In contrast to \cref{eq:boosting-constraint-rh}, an equality constraint is enforced, as heat pump operation is only permitted in conjunction with \ac{PTES} discharge.

\begin{subequations}
\begin{align}
 & \alphaHpThIn_{n,t} =
    \begin{cases}
      \dfrac{\retTemp_n - \bottomTemp_n}{\topTemp_n - \bottomTemp_n} &\!, \topTemp_{n} < \fwdTemp_n\\
      0 & \!, \topTemp_n \geq \fwdTemp_{n,t}
   \end{cases} & \forall n \in \N, t \in \T \label{eq:boosting-ratio-hp}\\
  & \thInputHP_{n,t}  = \alphaHpThIn_{n,t} \, \discharge_{n,t} \label{eq:bhp} \quad& \forall n \in \N, t \in \T\\
  &\thInputHP_{n,t}, \discharge_{n,t} \in \mathbb{R}^+ & \forall n \in \N, t \in \T \label{eq:bhp-def}
\end{align}
\end{subequations}

It should be noted that our formulations of both boosting configurations only approximate real-world operation. Most importantly, we pre-determine storage temperatures, which would actually depend on storage operation. 
When boosting via heat pumps, we thus assume that temperatures in the bottom layer below the return temperature are maintained even without heat pump operation, which is not achievable in practice.

\begin{figure}[ht]
  \centering
  \includegraphics[width=1.0\linewidth]{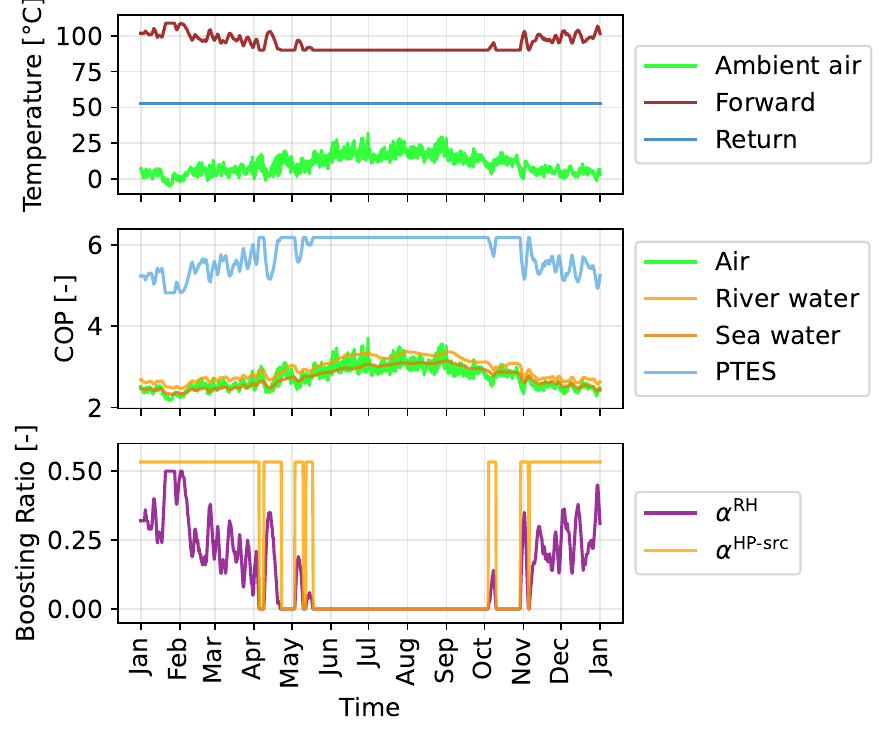}%
  \caption{District heating network temperatures, heat pump COPs and boosting ratios for resistive heaters and heat pumps exemplarily in the district heating system of Rostock for \textit{MidST} temperature levels, i.e. 108.8/52.7\textdegree C.}
  \label{fig:temperature_cops_boostingratios}
\end{figure}

\Cref{fig:temperature_cops_boostingratios} shows temperature levels and \acp{COP} for different heat sources alongside $\alphaRH$ and $\alphaRet$ for the Rostock district heating system assuming maximum forward/return temperatures of 108.8/52.7\textdegree C (see \cref{sec:scenarios}).

\subsection{Scenarios}

\label{sec:scenarios}

We investigate how \ac{PTES} affects the German energy system under current and reduced district heating network temperatures, and how its operational role is shaped by temperature-related constraints. 
In order to do so, a set of counterfactual model runs was designed. Each scenario is evaluated \emph{with and without} \ac{PTES}, such that the incremental system effects of \ac{PTES} can be isolated.  
First, runs without \ac{PTES} are computed, then investments in all countries but Germany are fixed and a second run with enabled \ac{PTES} investments for Germany is conducted, while no additional investments outside Germany are admitted.

Four \ac{PTES} boosting configurations are assessed: 
\begin{itemize}
  \item \noPTES does not allow any \ac{PTES} investments.
  \item \freeBoosting (unphysical) allows \ac{PTES} but ignores boosting requirements with $\bottomTemp=\retTemp$.
  \item \freeCapacity (unphysical) requires \ac{PTES} boosting via resistive heaters with $\bottomTemp=10$\textdegree C.
  \item \resistiveBoosting allows boosting via resistive heaters with $\bottomTemp=\retTemp$ (see \cref{sec:resistive-boosting}).
  \item \hpBoosting applies boosting via heat pumps with $\bottomTemp=10$\textdegree C (see \cref{sec:hp-boosting}). 
\end{itemize}

Under \freeCapacity, energetic \ac{PTES} capacities correspond to those of \hpBoosting to isolate the effect of increased storage capacity without requiring heat-pump investments, as introduced in \cref{sec:energetic-capacity}.
Both \freeCapacity and \freeBoosting are idealized reference cases that remove physical limits on storage boosting and energy capacity. While not technically feasible, these scenarios help isolate the mechanisms by which network temperature, boosting requirements, and energy capacity constraints influence the value and flexibility of \ac{PTES}.

Further, for each boosting configuration, three district heating supply temperature levels are investigated: 
\begin{itemize}
  \item \highTemp: today's forward and return temperatures\\(124/60\textdegree C)
  \item \midTemp: annual temperature reductions of 0.5\% until 2045 (108.8/52.7\textdegree C)
  \item \lowTemp: annual temperature reductions of 1\% until 2045 (95.5/46.2\textdegree C)
\end{itemize}
For all temperature levels, we assume constant \ac{PTES} top temperatures of 90\textdegree C and admit no forward temperatures below that value.

Thus, 15 scenarios are computed in total, the results of which are presented in the following section.

\section{Results}
\label{sec:results}

Our analysis starts with a comparison of a counterfactual case without \ac{PTES} and a reference case with \ac{PTES} and resistive boosting at \midTemp to illustrate the operational impacts of \ac{PTES} integration. 
In particular, \cref{sec:results_area_plot} examines temporally resolved and \cref{sec:results_single_systems} spatially resolved district heat balances and prices. 
\Cref{sec:results_system_savings} quantifies the effect of \ac{PTES} on the total system costs across all scenarios, and \cref{sec:ptes_price_temp_sensitivity} explores the response of \ac{PTES} discharge to electricity prices and network temperatures under different boosting regimes.

\subsection{Operational integration of \ac{PTES} in district heating}
\label{sec:results_area_plot}

\begin{figure*}[!t]
    \centering
    \includegraphics[width=\textwidth]{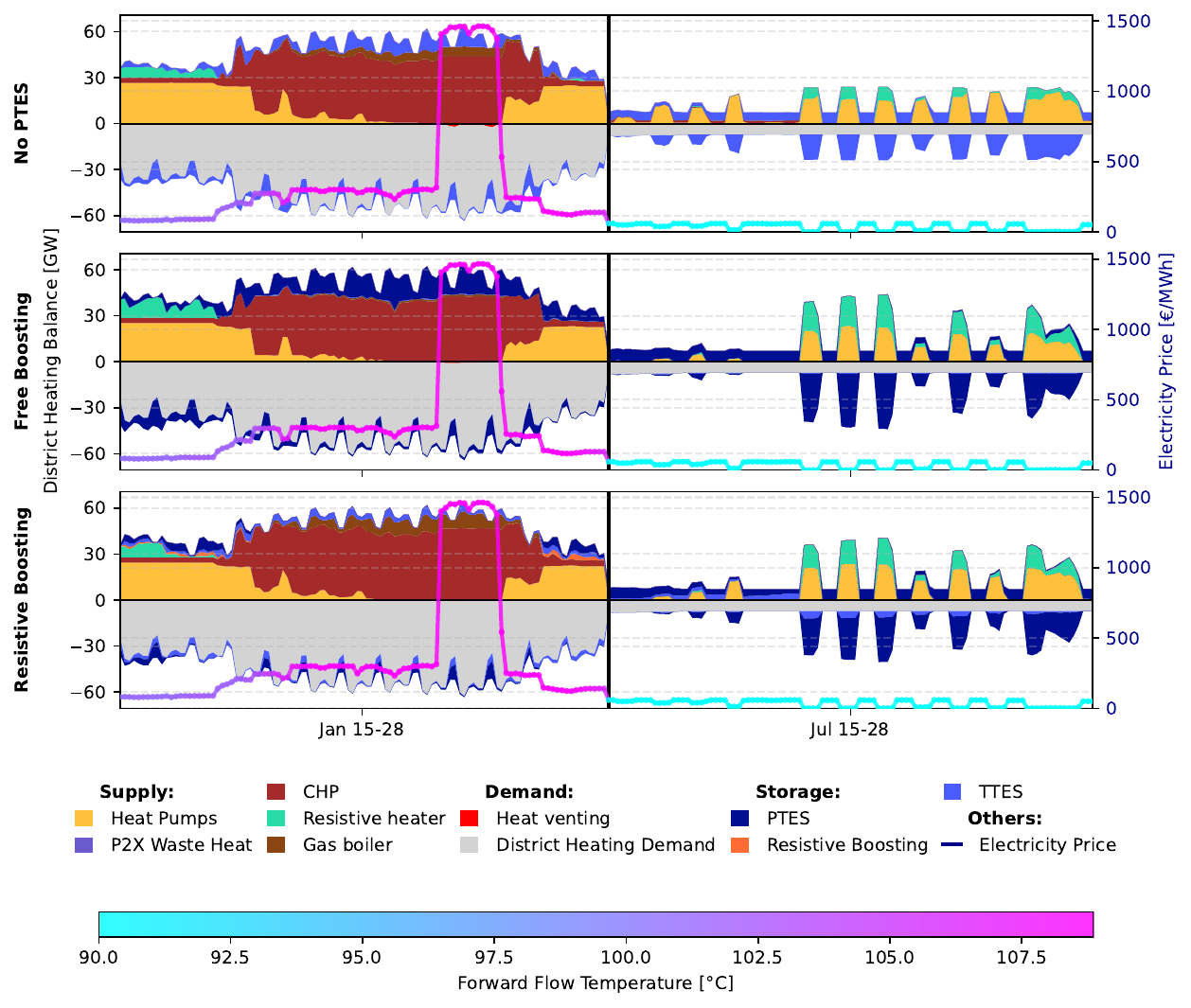}
    \caption{Spatially aggregated district heating balances in Germany for two example weeks in January and July 2045 at \midTemp. The top panel shows results without \ac{PTES}, the middle panel with \ac{PTES} and free boosting (unphysical scenario for reference), and the bottom panel with \ac{PTES} and resistive boosting. The secondary y-axis shows the average electricity price and the color of the price curve the corresponding average forward temperature in district heating networks.}
    \label{fig:dh_balance_aggregated}
\end{figure*}

\noindent
\Cref{fig:dh_balance_aggregated} shows the aggregated district heating dispatch in Germany for two example winter and summer weeks without as well as with \ac{PTES} under \freeBoosting and \resistiveBoosting and \midTemp together with electricity prices and forward temperatures.
The winter week begins with low heat demand, forward temperatures around 100\,\si{\celsius}, and electricity prices below 90\,\euro{}/MWh. In this regime, district heating is covered almost entirely by power-to-heat, with heat pumps and resistive heaters charging thermal storage at night and discharging during daytime peaks. As a cold spell develops, forward temperatures rise towards 108\,\si{\celsius} and heat demand exceeds 60\,\si{\giga\watt}. The cold spell coincides with a \emph{Dunkelflaute}, which is reflected in spiking electricity prices. With rising electricity prices, resistive heaters are curtailed first. Heat pump output declines and then ceases at price extremes. Fuel-fired \ac{CHP} plants then supply the bulk of district heat load, leveraging co-generation revenues in the power market, while gas boilers cover remaining peaks.

\ac{PTES} and \ac{TTES} are operated flexibly across all scenarios, with charging predominantly at night and discharging during daytime demand peaks. Aggregation across systems reveals asynchronous operation, visible as simultaneous charge and discharge bands in \cref{fig:dh_balance_aggregated}. With \ac{PTES} boosted by resistive heaters, the temperature gap between storage and network necessitates auxiliary energy for boosting, visible as the orange band in the middle panel. The loss associated with boosting is negligible under the moderate conditions at the beginning of the week because standalone power-to-heat is also part of the optimal dispatch in the counterfactual scenarios without \ac{PTES} and with idealized \ac{PTES} that does not require any boosting. During the cold \emph{Dunkelflaute}, however, the additional energy required for boosting limits \ac{PTES} discharge; by contrast, \ac{TTES} and the idealized \ac{PTES} case without boosting constraints sustain higher discharge capacity factors and act as the principal complements to \ac{CHP} at demand peaks. \ac{TTES} discharge is not directly affected by the boosting required for \ac{PTES} discharge. However, \ac{TTES} capacities are lower than in the \emph{No PTES} counterfactual scenario, leading to a lower contribution to supply. Overall, total \ac{TES} discharge during the cold spell is lower under \resistiveBoosting than in the system relying solely on \ac{TTES}, and the gap is filled by increased gas boiler dispatch. Even in these critical periods, \ac{PTES} adds value by providing high-capacity, flexible offtake for excess heat from \ac{CHP}; reduced heat venting during the \emph{Dunkelflaute} indicates this effect. Once electricity prices fall below about 150\,\euro{}/MWh, \ac{PTES} resumes discharge despite high forward temperatures, implying that resistive boosting remains cost-effective at these moderate price levels.

\medskip
\noindent
In the summer week, electricity prices fluctuate between 0 and 60\,\euro{}/MWh and forward temperatures are at the 90\,\si{\celsius} floor. A persistent low heat demand remains, mainly from low-tem\-per\-a\-ture in\-dus\-tri\-al uses. Price dynamics again shape operations: low-price hours trigger heat pump supply and storage charging. With \ac{PTES} present, charging becomes spikier as resistive heaters are dispatched; generation exceeds 40\,\si{\giga\watt} towards the end of the first week, more than five times the district heating load. Conversely, at night and when prices exceed about 55\,\euro{}/MWh (all nights except from Jul 26 to Jul 27), \ac{TTES} and \ac{PTES} provide most of the heat supply.
\Cref{fig:dh_balance_aggregated} shows temporally resolved \ac{PTES} behaviour under varying temperature and price conditions. To understand how these operational effects translate into structural changes across individual systems, the following section compares geographically resolved annual supply compositions and price impacts.

\subsection{Impact on district heating mix across systems}
\label{sec:results_single_systems}

\begin{figure*}[!t]
    \centering
    \includegraphics[width=\textwidth]{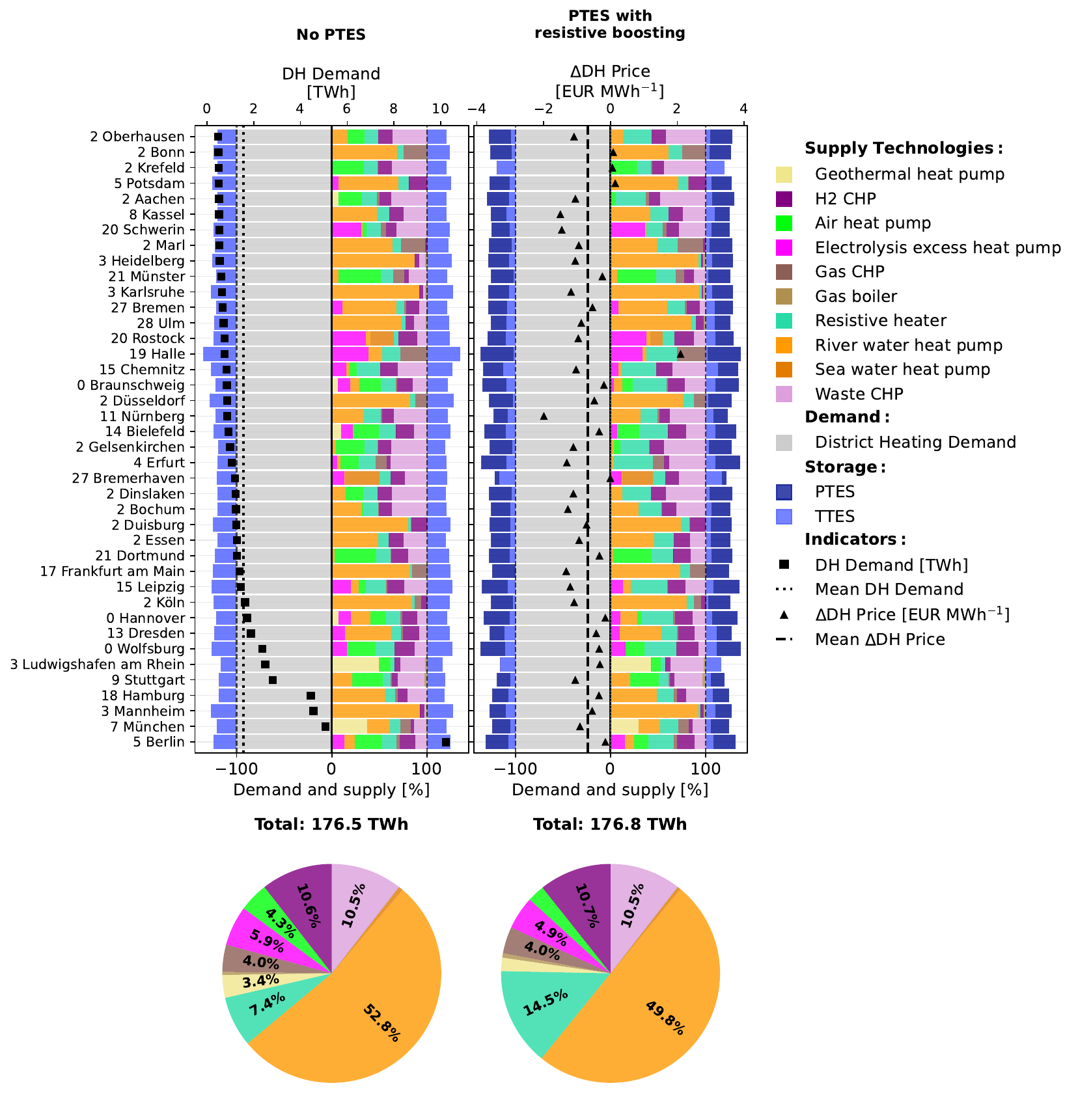}
    \caption{Temporally aggregated district heating balances in Germany for 40 largest systems at \midTemp in 2045. The left panel shows results without \ac{PTES} and the right panel with \ac{PTES} boosted by resistive heaters. The secondary x-axes show the system-specific and average annual district heating demands (left) and the system-specific and average savings in district heating prices (right). On the bottom the aggregated supply mix is shown including the base regions.}
    \label{fig:dh_system_balance_MT_rhboost}
\end{figure*}

Temporally aggregated balances for the 40 largest German district heating systems \emph{without PTES} and under \resistiveBoosting in the \midTemp case are shown in \cref{fig:dh_system_balance_MT_rhboost} (analogous results for \hpBoosting are provided in \cref{fig:single_dh_systems_energy_balance_hpboost}). The plot highlights pronounced heterogeneity in supply mixes arising from locally variable availability of renewable heat sources, \ac{PTES} siting potentials, electricity price profiles, and brownfield \ac{CHP} capacities.

\Ac{PTES} is widely adopted despite this diversity with the exceptions of Krefeld and Ludwigshafen am Rhein due to absent potentials. The complementary operation with \ac{TTES} observed in \cref{sec:results_area_plot} is confirmed at single-system level. With \ac{PTES} being available, all systems still retain \ac{TTES} except Frankfurt am Main. \Ac{CHP} generation is present in every system, so none relies exclusively on electrified supply. With or without \ac{PTES}, non-electrified (predominantly \ac{CHP}-fired) heat generation supplies about one quarter of German district heating, implying that for \resistiveBoosting under \midTemp, additional flexibility from \ac{PTES} does not, by itself, raise overall electrification compared to a system with \ac{TTES} only.
Despite the complementary role of fuel-fired supply, electrified technologies dominate overall generation. River-water potentials are fully exhausted in all systems with access and provide roughly half of total German district heating generation. Geothermal potentials, however, reach substantial utilisation only in Ludwigshafen am Rhein (100\% without and 87\% with \ac{PTES}) and Munich (54\% with and 43\% without \ac{PTES}), while their contribution remains small in all other systems.

Across systems, \ac{TES} discharge supplies 45\,\si{TWh} in total (12\,\si{TWh} \ac{TTES}, 33\,\si{TWh} \ac{PTES}); without \ac{PTES}, \ac{TTES} discharge reaches 39\,\si{TWh}, evidencing flexibility gains with \ac{PTES}. The mix among electrified supply technologies changes: resistive heater contribution rises by 7\%, while heat pump contributions fall by 6\% in total (roughly -3\% \ac{RWSHP}, -2\% \ac{ASHP}, -1\% geothermal). The cumulative charged-energy curve (\cref{fig:ptes_charge_cumulative}) shows strong concentration (50\% of charged energy in 3\% of hours), implying low charging capacity factors.
These low capacity factors alongside the much higher specific \ac{CAPEX} of heat pumps relative to resistive heaters (see \cref{tab:capex}) favors resistive charging.

Additional flexibility and the resulting supply shift reduce demand-weighted district heating prices by 0.7\,\euro{}/MWh on average in absolute terms (ranging from $-2.3$\,\euro{}/MWh in Halle to $+2$\,\euro{}/MWh in Nürnberg) and by 1.2\% on average in relative terms (from $-5$\% in Halle to 3.3\% in Nürnberg) when \ac{PTES} with resistive boosting is added.

\subsection{Impacts of network temperatures on the value of \ac{PTES}}
\label{sec:results_system_savings}

\begin{figure*}[!t]
    \centering
    \includegraphics[width=\textwidth]{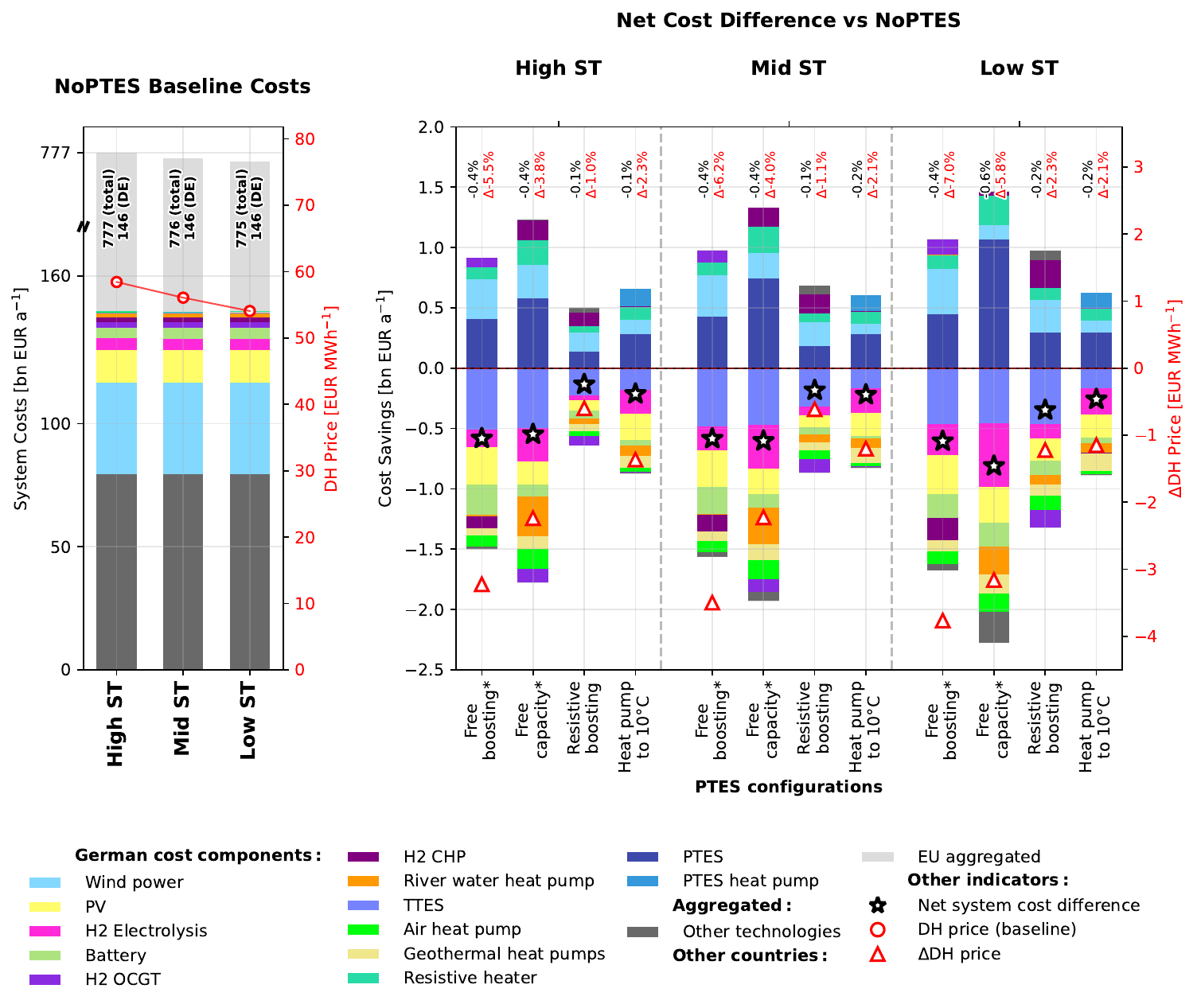}
    \caption{Total system costs by technology (\ac{CAPEX} and \ac{OPEX} aggregated) for different supply temperature levels without PTES (left) and with PTES (right) as difference. The total savings when PTES is introduced to the system are shown for the different boosting scenarios. Additionally, the demand-weighted average district heating prices are shown for each of the scenarios as absolute value (NoPTES) and difference (with PTES). While German costs are distinguished, the costs incurred in EU neighbor countries are aggregated. Furthermore, technologies not affected by the introduction of PTES are aggregated.\\ * The idealized unphysical scenarios are marked with asterisks.}
    \label{fig:system_savings}
\end{figure*}

The preceding analysis centered on operational integration at \midTemp with resistive boosting to establish core interactions between \ac{PTES} and district heating. We now extend to all three temperature scenarios, and all four \ac{PTES} configurations to assess system-wide impacts. 
The left-hand side of \cref{fig:system_savings} shows total system costs by technology alongside average district heating prices \emph{without PTES} as reference cases. In these reference cases, we observe a substantial decrease of district heating prices at reduced network temperatures  without \ac{PTES}. From \highTemp to \lowTemp they decrease by 8\%. The right-hand side shows the cost differences relative to these reference cases when introducing \ac{PTES} under idealised conditions with free boosting and free capacity, as well as under feasible boosting configurations using resistive or heat-pump boosting.

In all feasible scenarios, total system costs and average district heating prices decline with \ac{PTES}. Savings reach 134\,M\euro{}\,a$^{-1}$ (\resistiveBoosting) and 213\,M\euro{}\,a$^{-1}$ (\hpBoosting) at \highTemp, with price reductions of 0.60 and 1.36\,\euro{}/MWh; 183 and 220\,M\euro{}\,a$^{-1}$ under \midTemp (0.61 and 1.20\,\euro{}/MWh); and 347 and 258\,M\euro{}\,a$^{-1}$ at \lowTemp (1.22 and 1.14\,\euro{}/MWh). Through the joint impact of network temperature reductions together with \ac{PTES}, district heating prices are reduced by up to 10\%.

The idealized scenarios clarify the impact of the temperature gap between the \ac{PTES} top temperature and the forward flow in district heating networks, as well as the gap between the return flow and the bottom storage temperature that is theoretically achievable with heat-pump cooling.
Relative to the \emph{No PTES} scenarios, the discrepancy in German system savings range from approximately 0.2-0.4\%; for district heating prices, the gap spans from 1-5\%.

The difference between \resistiveBoosting and \freeBoosting narrows as temperatures fall, indicating that boosting demand decreases with lower network temperatures. At \lowTemp, the additional savings from \freeBoosting still reach 258~M\euro{}\,a$^{-1}$, which could be realized by further reducing forward temperatures from 95.5°C to 90°C.
The persistent disparity in district heating price reductions between resistive and free boosting shows that boosting remains a major obstacle to deeper price reductions in district heating systems that adapt \ac{PTES} even for small temperature gaps.

The cost gap between \resistiveBoosting and the idealized \freeCapacity case increases further at \lowTemp (467\,M\euro{}\,a$^{-1}$), underscoring the limitation that the return temperature imposes on the usable energy capacity in \ac{PTES}. The design of the idealized scenarios equalizes volumetric capacity between resistive and heat-pump scenarios confering an advantage that shrinks with falling return temperatures (since additional capacity declines), implicitly assuming larger reductions of \ac{PTES} CAPEX and boosting costs at higher temperatures. As the system cost gap still remains largest at \lowTemp, the system value of \ac{PTES} energy capacity rises with declining network temperatures.

The cost gaps to the idealised scenarios highlight the additional savings that could be unlocked at even lower network temperatures below the scope of our scenarios, where reduced forward temperatures would limit boosting requirements and lower return temperatures would expand usable storage capacity. Even so, the feasible configurations already deliver consistent system savings across all temperature levels, even at current conditions in Germany. This shows that \ac{PTES} provides system value even if temperature reductions in district heating networks remain limited.

\subsection{Influence of boosting technologies across temperature levels}
\label{sec:results_boosting_technologies}

Within each temperature scenario, the savings differ between resistive boosting and booster heat pumps. Furthermore, the two configurations respond differently to temperature changes.
Heat pumps are favored at \highTemp and \midTemp, while resistive heaters yield higher savings at \lowTemp. The heat pump advantage at higher temperatures follows from higher efficiency: less electricity is required to enable \ac{PTES} discharge, reducing the additional power load when \ac{VRES} are scarce, as it often is the case during winter, particularly during cold \emph{Dunkelflaute} events. Another reason for the stronger heat-pump performance at higher storage temperatures is the larger usable energy capacity. It enables more charging during periods of low electricity prices increasing the share of low-cost energy stored in \ac{PTES} and further improving overall savings. In contrast, the higher reliance on charging \ac{PTES} at higher electricity prices with \resistiveBoosting is confirmed by the patterns in \cref{fig:ptes_discharge_charge_by_price_and_T} and the higher cycles shown in \cref{fig:ptes_soc_comparison_distribution}.

The growing advantage of resistive boosting at low supply temperatures arises from three effects: 

(i) lower forward temperatures reduce the boosting ratio, thereby decreasing the auxiliary electricity required for \ac{PTES} discharge; 
(ii) lower return temperatures widen the \ac{PTES} operating temperature range, which increases its usable energy capacity; 
and (iii) the resulting larger usable energy capacity further lowers boosting costs, because the ratio of auxiliary boosting energy to \ac{PTES} discharge declines.

For booster heat pumps, temperature reductions impose opposing effects:
(i) lower source temperatures reduce average \ac{COP} (from 5.5 to 4.9 from \highTemp to \lowTemp);
(ii) while more direct use of \ac{PTES} discharge slightly reduces required heat pump capacity (from 11.1 to 10.8\,\si{\giga\watt_{th}}).
Rising system savings indicate, that the latter effect dominates.

\subsection{Change of investment patterns across sectors}
\label{sec:results_investment_patterns}

The net cost difference achieved through \ac{PTES} integration arise from changes in both investment and operational decisions across multiple technologies. These shifts mirror the altered district heating operation described in \cref{sec:results_area_plot} and \cref{sec:results_single_systems}, but they also extend beyond the boundaries of the district heating system, as shown in \cref{fig:system_savings}.

The optimisation model achieves these savings by \ac{PTES} investments of 0.9\,\si{TWh} (27\,\si{\mega\meter\cubed}) with \resistiveBoosting and 5\,\si{TWh} (53\,\si{\mega\meter\cubed}) under \hpBoosting under \highTemp; 1.5\,\si{TWh} (35\,\si{\mega\meter\cubed}) and 5\,\si{TWh} (54\,\si{\mega\meter\cubed}) at \midTemp; and 2.9\,\si{TWh} (56\,\si{\mega\meter\cubed}) and up to 5.3\,\si{TWh} (57\,\si{\mega\meter\cubed}) at \lowTemp.
The discrepancies in the energy-capacity-to-volume ratio are due to the different temperature spreads that each boosting configuration can exploit.

Booster heat-pump investments amount to 120–140 M€ a$^{-1}$ from low to high temperatures, corresponding to 40–50\% of \ac{PTES} investments.
Owing to their lower \ac{CAPEX}, resistive-heater capacities rise in all configurations to charge \ac{PTES} by up to 26\% compared to runs without \ac{PTES}, whereas total stand-alone heat-pump capacity decreases by up to 2\%.
Resistive boosting does not entail additional resistive-heater investments relative to heat-pump boosting. Investments are in fact slightly higher under heat-pump boosting, as cooling of the storage bottom temperature increases the usable energy capacity of \ac{PTES}, allowing more extensive exploitation of low-price periods for resistive charging.

Irrespective of temperature levels, under \hpBoosting, 0.6--0.7\,\si{TWh} of \ac{TTES} is retained to complement \ac{PTES}; with \resistiveBoosting, \ac{TTES} falls from 0.5\,\si{TWh} (\highTemp) to 0.4\,\si{TWh} (\midTemp) and 0.04\,\si{TWh} (\lowTemp). 
Furthermore, additional hy\-dro\-gen-fired \ac{CHP} capacity is built with \resistiveBoosting compared to scenarios without \ac{PTES} to ensure flexible supply during periods of high prices and high forward temperatures (+4, +6, and +8\,\si{\giga\watt_{th}} at \highTemp, \midTemp, and \lowTemp). Because these plants also serve as peak power generators, their investments are partially offset by lower requirements for hydrogen \ac{CCGT} peakers and batteries. In \hpBoosting scenarios, a modest shift from PV–battery pairing toward more wind (up to +2\%) also contributes to reduced battery needs. Further evidence of added flexibility is the reduction in electrolyzer and hydrogen storage capacity in all scenarios (both up to \(-4\%\) at \lowTemp with booster heat pumps), though absolute cost effects remain small due to low specific storage CAPEX (1.46\,\euro{}/kWh).

 \subsection{\ac{PTES} dispatch across electricity prices and network temperatures}
\label{sec:ptes_price_temp_sensitivity}

\begin{figure*}[!t]
    \centering
    \includegraphics[width=\textwidth]{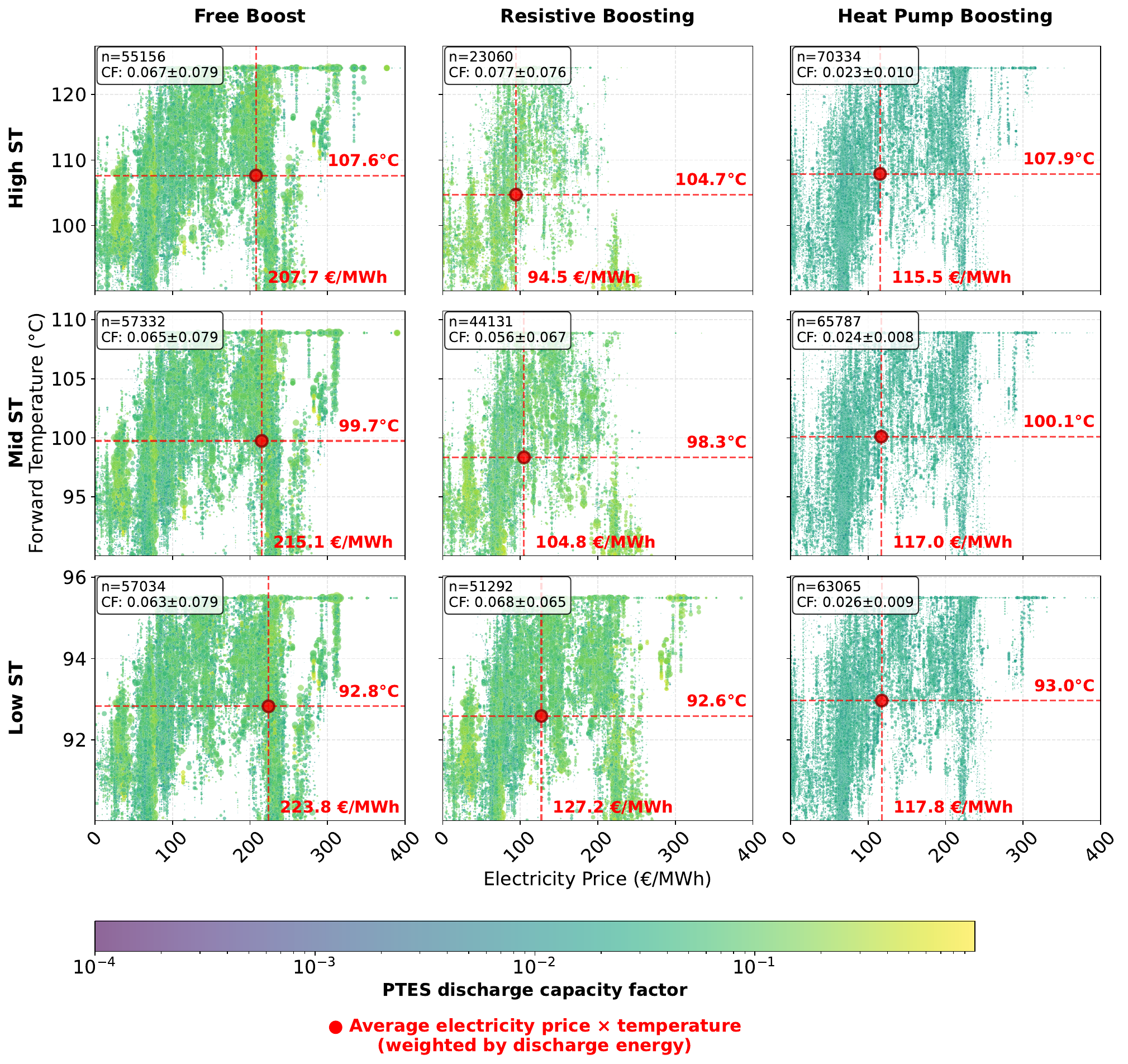}
    \caption{Discharge events of \ac{PTES} in all district heating systems over electricity prices and district heating network forward temperatures at times where forward temperature is above 90\,\si{\celsius} and boosting is required. The discharge distribution is shown for all three network temperature scenarios, both feasible boosting configurations, as well as the idealized scenario where no boosting is required. Color indicates the capacity factor of the \ac{PTES} discharge, and marker size the discharged energy in each timestep. The red marker shows the average electricity price and forward temperature during boosted discharge events.}
    \label{fig:ptes_discharge_sensitivity}
\end{figure*}

We complement the aggregated metrics of \cref{sec:results_area_plot,sec:results_single_systems,sec:results_system_savings} with an analysis of how \ac{PTES} discharge responds to electricity prices and forward temperatures.

\Cref{fig:ptes_discharge_sensitivity} shows how boosting requirements shape \ac{PTES} operation across network temperature scenarios: Each dot represents a discharge event at a combination of forward temperature and electricity price in the respective district heating system or model region for different temperature scenarios under \resistiveBoosting, \hpBoosting and \freeBoosting. The dots are coloured by the amount of discharge energy relative to the maximum wattage, i.e. the respective capacity factor.

With \resistiveBoosting, the average electricity price at which boosted discharge occurs increases from 94.5\,\euro{}/MWh (\highTemp) to 104.8\,\euro{}/MWh (\midTemp) and 127.2\,\euro{}/MWh (\lowTemp). Capacity factor dispersion is high (coefficient of variation 0.96–1.2) at low means (0.06–0.08), consistent with a volatile discharge pattern. 
With \hpBoosting, the corresponding average prices rise modestly from 115.5 to 117.0 and 117.8\,\euro{}/MWh, while capacity factors are lower and far less dispersed (coefficient of variation 0.03–0.04; means 0.02–0.03), indicating steadier operation.

These patterns reflect the efficiency advantage of heat pumps over resistive heaters: heat pumps support more frequent boosted discharge at higher prices and forward temperatures. Only at the lower forward temperatures of the \lowTemp scenario (below 96\,\si{\celsius}), resistive-boosted discharges become sufficiently inexpensive to participate even during high-electricity-price periods, including peaks above 275\,\euro{}/MWh. At such low network temperatures, discharge patterns converge towards the idealized free-boosting case, confirming the diminishing boosting costs highlighted in \cref{sec:results_system_savings}. Nevertheless, some high-price discharge events observed in the idealized \freeBoosting case (prices $>$250\,\euro{}/MWh at mid-to-low forward temperatures within the scenario ranges) remain uneconomic with either feasible configuration because alternative flexible tech\-no\-lo\-gies---notably hydrogen-fired \ac{CHP} and \ac{TTES}---are less costly at those points, as we point out in \cref{sec:results_area_plot,sec:results_single_systems}. During coincident cold spells and \emph{Dunkelflaute} events, when prices and forward temperatures are simultaneously high, boosted discharge still occurs; booster heat pumps can sustain discharge even up to 124\,\si{\celsius}, in line with current German systems.

The higher count of boosted discharge events alongside low, low-variance capacity factors with \hpBoosting points to an economic constraint induced by the additional investment: booster heat pumps operate only when forward temperatures exceed 90\,\si{\celsius} and thus run at high capacity factors during these periods to recover costs, largely independent of short-term demand and \ac{COP} fluctuations. This investment-driven constraint shapes \ac{PTES} into a steady, low-power winter supplier under heat pump boosting, whereas resistive boosting preserves more flexible discharge. These different operational characteristics are confirmed by the temporally resolved district heating balances shown in \cref{fig:dh_balance_aggregated_all_HT,fig:dh_balance_aggregated_all_MT,fig:dh_balance_aggregated_all_LT}.

\section{Discussion}
\label{sec:discussion}

\subsection{Comparison with literature}

Our findings on the role of \ac{PTES} are broadly consistent with the limited literature on large-scale thermal storage in sector-coupled systems, but also highlight important differences in scale and temperature dependence. Sifnaios et al.~\cite{sifnaios_impact_2023} remains, to our knowledge, the only study that explicitly assesses \ac{PTES} and \ac{TTES} in a multi-country large-scale model. For Denmark, they report around 3.8\,TWh of \ac{PTES} for 39.9\,TWh of district heating demand, implying a higher storage-to-demand ratio than in our German case, where up to 5.3\,TWh of \ac{PTES} serve about 150\,TWh of demand. This discrepancy is in line with deviating model assumptions, which comprise lower \ac{PTES} \ac{CAPEX} in their study, higher degree of aggregation, which inflates TES expansion, and lower network temperatures that do not require discharge boosting. Still, their scenario that combines \ac{TTES} and \ac{PTES} sees a complementary role for \ac{TTES}. In addition, our idealized free-capacity scenario confirms the high sensitivity of \ac{PTES} deployment and value to investment cost assumptions, in line with the strong \ac{CAPEX} sensitivity reported by Sifnaios et al.~\cite{sifnaios_impact_2023}.

Although they do not include \ac{PTES}, the studies by Kök et al.~\cite{kokAchievingClimateNeutrality2025} and Billerbeck et al.~\cite{billerbeck_integrating_2024} provide a useful benchmark for temperature effects on district heating systems. Kök et al.\ show that reducing maximum supply temperatures—from 95\,\si{\celsius} to 65\,\si{\celsius}—lowers district heating costs by about 20\%. In our framework, lowering maximum forward temperatures from 124\,\si{\celsius} to 95\,\si{\celsius} reduces costs by roughly 7.6\% without \ac{PTES} and by about 10\% when \ac{PTES} with \resistiveBoosting is included. This suggests a lower temperature sensitivity in the higher temperature range relevant for current German systems, while highlighting that more progressive temperature reductions could enable substantially more efficient integration of low-temperature heat sources.
Billerbeck et al.\cite{billerbeck_integrating_2024} find that heat pumps provide the dominant share of district heating supply across all of their scenarios. They also report that fuel-fired generation from \ac{CHP} gains importance at higher network temperatures, reaching about 20\% of supply in their high-temperature case. Our results show the same tendency: in the medium-temperature scenarios for Germany, \ac{CHP} plants supply roughly 25\% of district heating, down from today's 86\% \cite{agfw_hauptbericht}.

\subsection{Limitations}

The main limitations of this work relate to the representation of temperature, weather, and technology and cost assumptions.

First, temperature and storage physics are simplified. Forward temperatures are fixed by scenario assumptions, and limiting them to a minimum of 90\,\si{\celsius} during warm periods may underestimate the potential to integrate low-temperature renewable heat or operate heat pumps at higher \ac{COP}s. 
Conversely, assuming constant source and sink inlet temperatures for booster heat pumps can overestimate achievable \ac{COP}s.
The fixed temperature spread of 90–10\,\si{\celsius} further overestimates the usable energy capacity of \ac{PTES} with heat-pump boosting, particularly in summer when heat-pump operation is suspended. 
For \resistiveBoosting, the assumption of a constant 90\,\si{\celsius} top-layer temperature is plausible because the high number of cycles (16.3–27.3 a\textsuperscript{-1}) and the continuous charging shown in \cref{fig:ptes_soc_comparison_distribution,fig:ptes_discharge_charge_by_price_and_T} during the winter months provide frequent reheating opportunities. For booster heat pumps, the modeled annual cycling frequencies (6.6–6.9 a\textsuperscript{-1}) are close to those observed at the Høje-Taastrup pit storage (6.5 in 2023 and 8.2 in 2024 \cite{sifnaiosPerformanceAnalysisHoje2025, sifnaiosExperiencesFirstShortTerm2023}). However, \cref{fig:ptes_soc_comparison_distribution} shows only limited charging from the beginning of the year until February and again from November to year-end across all temperature levels, which may indicate a misalignment between the fixed-temperature assumption and the storage dynamics implied by the optimized operation.
Furthermore, \ac{PTES} is represented without stratification and temperature-dependent standing losses. In practice, storage systems might tolerate moderate increases in top temperatures up to about 95\,\si{\celsius} and cooling below 10\,\si{\celsius}, which would increase usable energy capacity and alter operational constraints \cite{danishenergyagencyTechnologyDataEnergy2025}. Despite these simplifications, our results provide meaningful insights into how \ac{PTES} contributes to net-zero energy systems and how temperature-dependent boosting requirements fundamentally shape this contribution.

Second, the analysis relies on a single weather year. Using one year of ambient temperatures and  \ac{VRES} availability—especially coincident cold spells and \emph{Dunkelflauten}—can under- or overestimate the required flexibility investments, particularly \ac{PTES}, since other years may show more extreme or more moderate conditions.

Third, technology portfolios and cost assumptions are necessarily simplified. Uniform \ac{CAPEX} and network temperature assumptions across district heating systems cannot capture local heterogeneity in heat pump options, geothermal conditions, or infrastructure constraints, nor do they reflect economies of scale that have proven particularly impactful in existing \ac{PTES} projects \cite{sifnaios_impact_2023,danishenergyagencyTechnologyDataEnergy2025}. Moreover, the aggregation of smaller district heating systems into larger base regions may artificially inflate the technical potential of renewable heat sources, which could further contribute to the strong role of river-water-sourced heat pumps in the model. Other potentially relevant heat sources and storage options, such as solar thermal, industrial waste heat, data centres, or aquifer storage, are excluded, and competition for urban land is not modelled. 

The results should therefore be interpreted as indicative of system-level interactions and trends rather than as a prescriptive design for individual systems.

\section{Conclusion}
\label{sec:conclusion}

This study assessed the role of \ac{PTES} in a national, sector-coupled net-zero energy system under current and reduced district heating temperature levels, explicitly modeling tem\-per\-a\-ture-de\-pen\-dent discharge boosting—a technical constraint that most large-scale energy system studies neglect. Even at today's high forward temperatures, \ac{PTES} provides measurable system value by absorbing low-price excess electricity and shifting heat supply away from periods with higher prices. Across temperature scenarios, annual German system cost savings range from 135 to 345\,M\euro{}\,a$^{-1}$, and district heating prices fall by up to 4\,\euro{}/MWh —relative to a counterfactual system relying exclusively on tank thermal energy storage (\ac{TTES}).

However, within the temperature ranges considered, \ac{PTES} remains structurally limited: because electrically boosted discharge depends on electricity prices, it cannot operate economically during the most severe scarcity periods when both forward temperatures and prices peak. These effects have been neglected in the literature to date. During such hours, the system relies primarily on \ac{CHP}-fired generation and \ac{TTES}.

Booster heat pumps and resistive heaters mediate these limitations in different ways. Booster heat pumps achieve high efficiencies even at elevated forward temperatures, enabling \ac{PTES} to discharge during periods with both high temperature requirements and comparatively high electricity prices. Their high specific \ac{CAPEX}, however, induce an operating pattern in which \ac{PTES} discharges at low but relatively constant capacity factors throughout winter, largely decoupled from short-term price fluctuations. In contrast, resistive boosting is available at low \ac{CAPEX} but is OPEX-driven, as it consumes more electricity due to its lower efficiency. Its operation is therefore highly sensitive to electricity prices and to the forward temperature.

A timely roll-out of \ac{PTES} should be supported, as it delivers measurable system-level savings and lowers district heating costs even under today’s high network temperatures in Germany. At the same time, integration costs at stagnant temperature levels remain high, reflecting technical and operational challenges common to all low-temperature renewable heat supply options when confronted with elevated network temperatures. Taken together, these conditions strengthen the case for policies that facilitate temperature reductions in district heating networks.

Future research should explore non-electrified boosting options, such as fuel-fired boilers or absorption heat pumps, which may alter the trade-offs between \ac{CAPEX}, efficiency, and operational flexibility identified in this study. Additionally, a more detailed analysis of the interaction between \ac{PTES} and other long-duration energy storage technologies—particularly hydrogen cavern storage, \ac{ATES}, and \ac{BTES}—would help clarify their roles as competitors or complements in providing flexibility across the electricity and heating sectors.

\section*{Acknowledgements}

We gratefully acknowledge funding by the German Federal Ministry for Economic Affairs and Climate Action (BMWK) through the project \textit{SysGF} under grant no. 03EI1075B.

\section*{Author Contributions}

\textbf{Caspar Schauß}:
Conceptualization --
Data curation --
Formal Analysis --
Investigation --
Methodology --
Software --
Visualization --
Writing - original draft --
Writing - review \& editing

\textbf{Amos Schledorn}:
Conceptualization --
Methodology --
Software --
Supervision --
Writing - original draft --
Writing - review \& editing

\textbf{Tom Kähler}:
Methodology --
Software --
Writing - review \& editing

\textbf{Kristina Schumacher}:
Methodology --
Writing - review \& editing

\textbf{Mathias Ammon}:
Conceptualization --
Methodology --
Writing - review \& editing

\textbf{Tom Brown}:

Conceptualization --
Supervision --
Writing - review \& editing

\section*{Declaration of Generative AI and AI-assisted technologies in the manuscript preparation process}
During the preparation of this work, the author(s) used generative AI tools (GPT, Claude Sonnet, and Claude Opus) to improve wording and assist with code development and refinement. After using these tools, the author(s) reviewed and edited all content as necessary and take full responsibility for the content of the published article.

\section*{Declaration of Competing Interest}
Kristina Schumacher and Matthias Ammon are professionally active as consultants in the district heating sector. No specific financial interests related to this study exist.

\section*{Data and Code Availability}

A dataset of the model results and the code to reproduce the experiments will be made available at \textit{Zenodo} and \textit{GitHub}, respectively, upon final publication.

\addcontentsline{toc}{section}{References}
\renewcommand{\ttdefault}{\sfdefault}
\bibliography{references}

\newpage

\renewcommand{\citenumfont}[1]{S#1}
\setcounter{equation}{0}
\setcounter{figure}{0}
\setcounter{table}{0}
\setcounter{section}{0}

\section{Supplementary Information}
\label{sec:si}

\begin{table}[htbp]
    \centering
    \footnotesize
    \caption{Investment costs for selected district heating technologies. Based on \cite{danishenergyagencyTechnologyDataEnergy2025} and expert opinion from \textit{HIR Hamburg Research Institute gGmbH}.}
    \begin{tabular}{l|c|l|l}
\toprule
 & Cost & Unit \\
\midrule
PTES & 68 & EUR/m$^3$ \\
TTES & 2.41 & EUR/kWh \\
Air-sourced heat pumps & 906.10 & EUR/kW$_{th}$ \\
River-water-sourced heat pumps & 779.24 & EUR/kW$_{th}$ \\
Sea-water-sourced heat pumps & 779.24 & EUR/kW$_{th}$ \\
Excess-heat-sourced heat pumps & 688.64 & EUR/kW$_{th}$ \\
Geothermal heat pumps & 598.03 & EUR/kW$_{th}$ \\
PTES heat pumps & 598.03 & EUR/kW$_{th}$ \\
\bottomrule
\end{tabular}

    \label{tab:capex}
\end{table}

\begin{figure*}[!t]
    \centering
    \includegraphics[width=\textwidth]{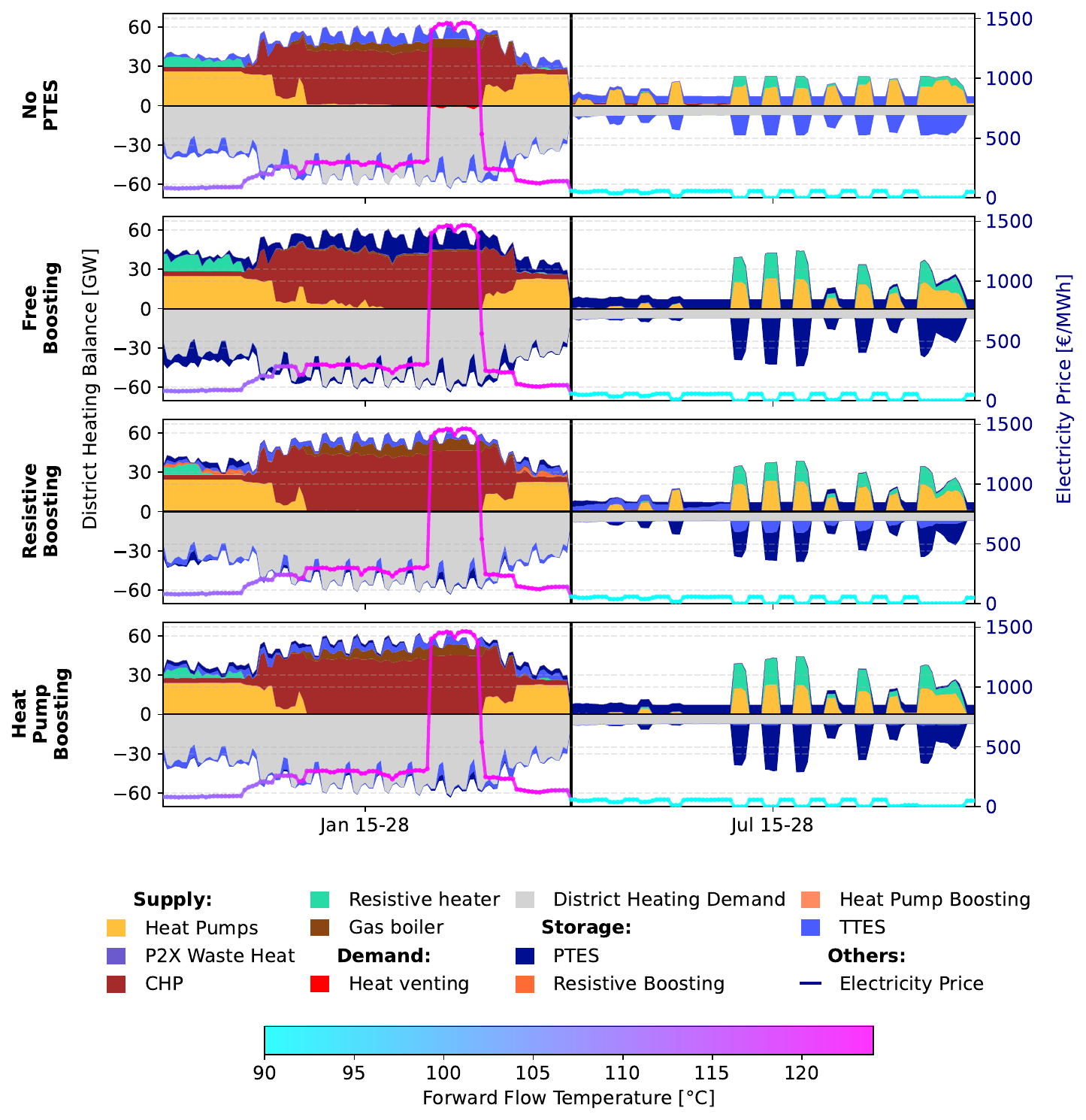}
    \caption{Spatially aggregated district heating balances in Germany for two example weeks in January and July at \highTemp. The top panel shows results without \ac{PTES}, the upper middle panel with \ac{PTES} and free boosting (unphysical scenario for reference), the lower middle panel with \ac{PTES} and resistive boosting, and the bottom panel with \ac{PTES} and heat pump boosting. The secondary y-axis shows the average electricity price and the corresponding average forward temperature in district heating networks.}
    \label{fig:dh_balance_aggregated_all_HT}
\end{figure*}

\begin{figure*}[!t]
    \centering
    \includegraphics[width=\textwidth]{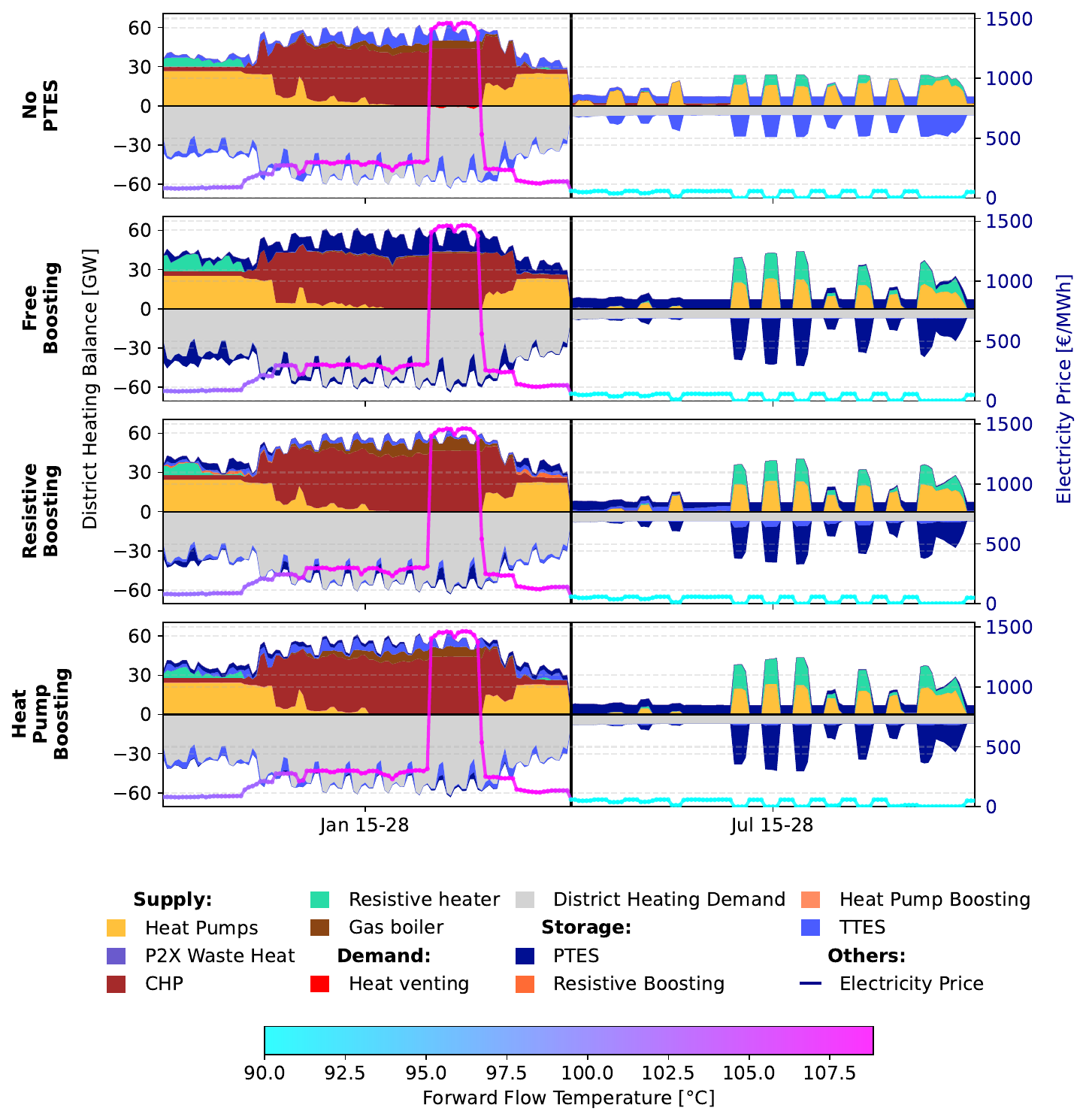}
    \caption{Spatially aggregated district heating balances in Germany for two example weeks in January and July at \midTemp. The top panel shows results without \ac{PTES}, the upper middle panel with \ac{PTES} and free boosting (unphysical scenario for reference), the lower middle panel with \ac{PTES} and resistive boosting, and the bottom panel with \ac{PTES} and heat pump boosting. The secondary y-axis shows the average electricity price and the corresponding average forward temperature in district heating networks.}
    \label{fig:dh_balance_aggregated_all_MT}
\end{figure*}

\begin{figure*}[!t]
    \centering
    \includegraphics[width=\textwidth]{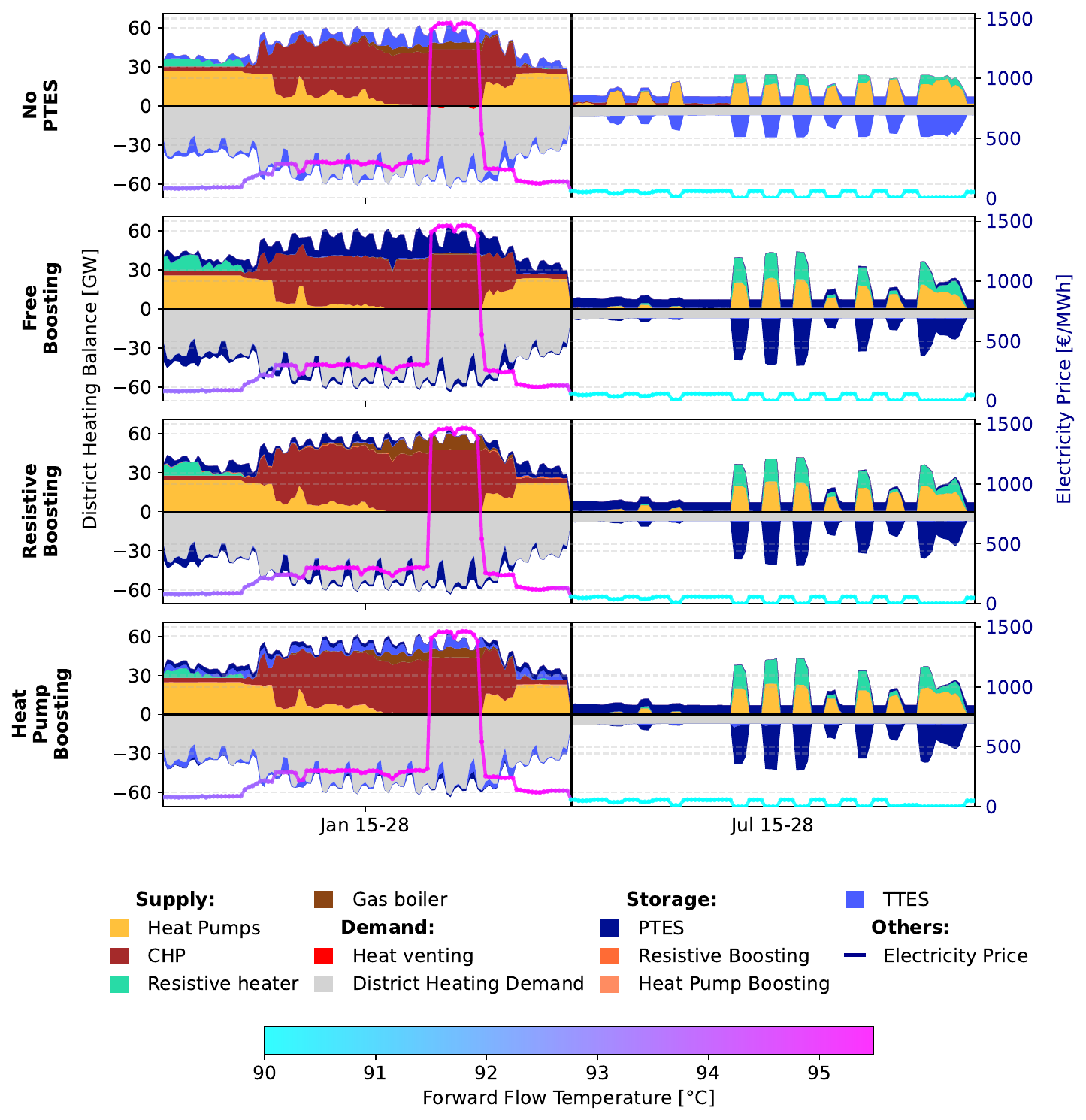}
    \caption{Spatially aggregated district heating balances in Germany for two example weeks in January and July at \lowTemp. The top panel shows results without \ac{PTES}, the upper middle panel with \ac{PTES} and free boosting (unphysical scenario for reference), the lower middle panel with \ac{PTES} and resistive boosting, and the bottom panel with \ac{PTES} and heat pump boosting. The secondary y-axis shows the average electricity price and the corresponding average forward temperature in district heating networks.}
    \label{fig:dh_balance_aggregated_all_LT}
\end{figure*}

\begin{figure*}[htbp]
    \centering
    \includegraphics[width=0.5\textwidth]{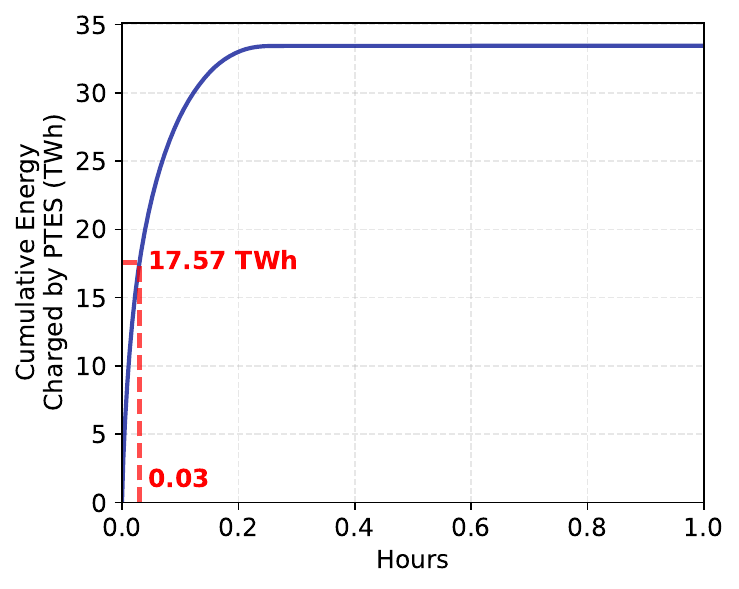}
    \caption{Cumulated charged energy in \ac{PTES} systems over all systems and the year, sorted by charge power. The curve corresponds to a scenario at medium network temperatures and an integration of \ac{PTES}with resistive boosting.}
    \label{fig:ptes_charge_cumulative}
\end{figure*}

\begin{figure*}[htbp]
    \centering
    \includegraphics[width=\textwidth]{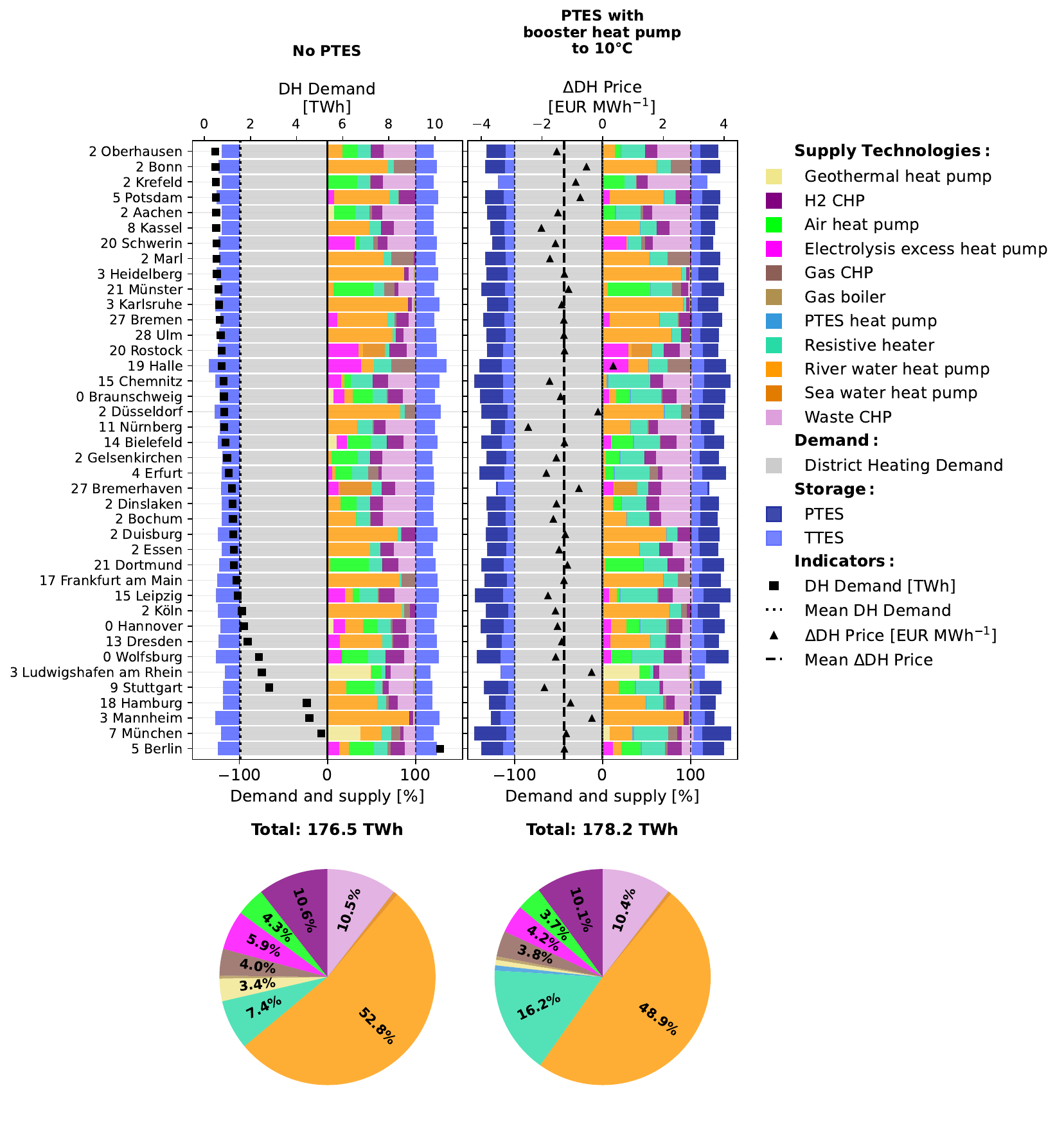}
    \caption{Temporally aggregated district heating balances in Germany for 40 largest systems at \midTemp. The left panel shows results without \ac{PTES} and the right panel with \ac{PTES} boosted by heat pumps. The secondary x-axes show the system-specific and average annual district heating demands (left) and the system-specific and average savings in district heating prices (right). On the bottom the aggregated supply mix is shown including the base regions.}
    \label{fig:single_dh_systems_energy_balance_hpboost}
\end{figure*}

\begin{figure*}[!t]
    \centering
    \begin{minipage}[t]{0.48\textwidth}
        \centering
        \includegraphics[width=\textwidth]{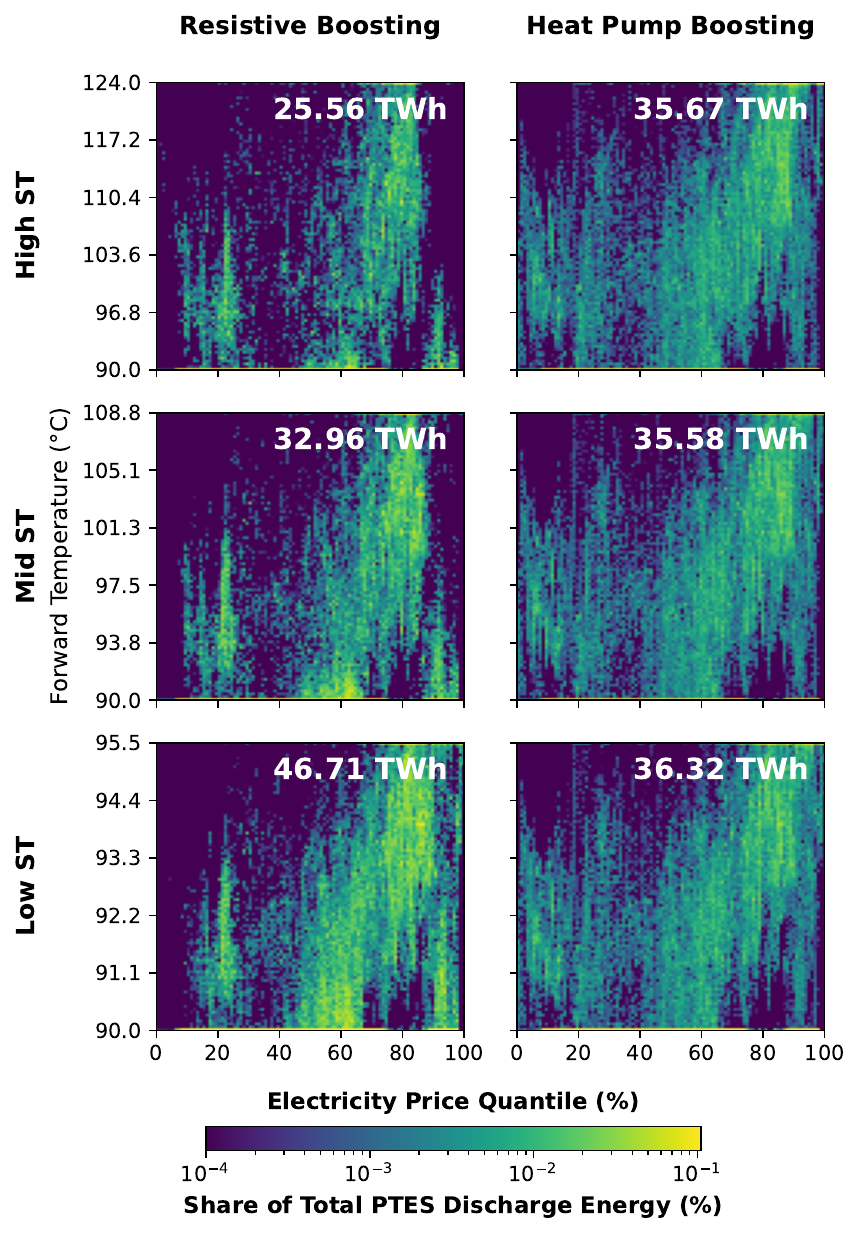}
    \end{minipage}
    \hfill
    \begin{minipage}[t]{0.48\textwidth}
        \centering
        \includegraphics[width=\textwidth]{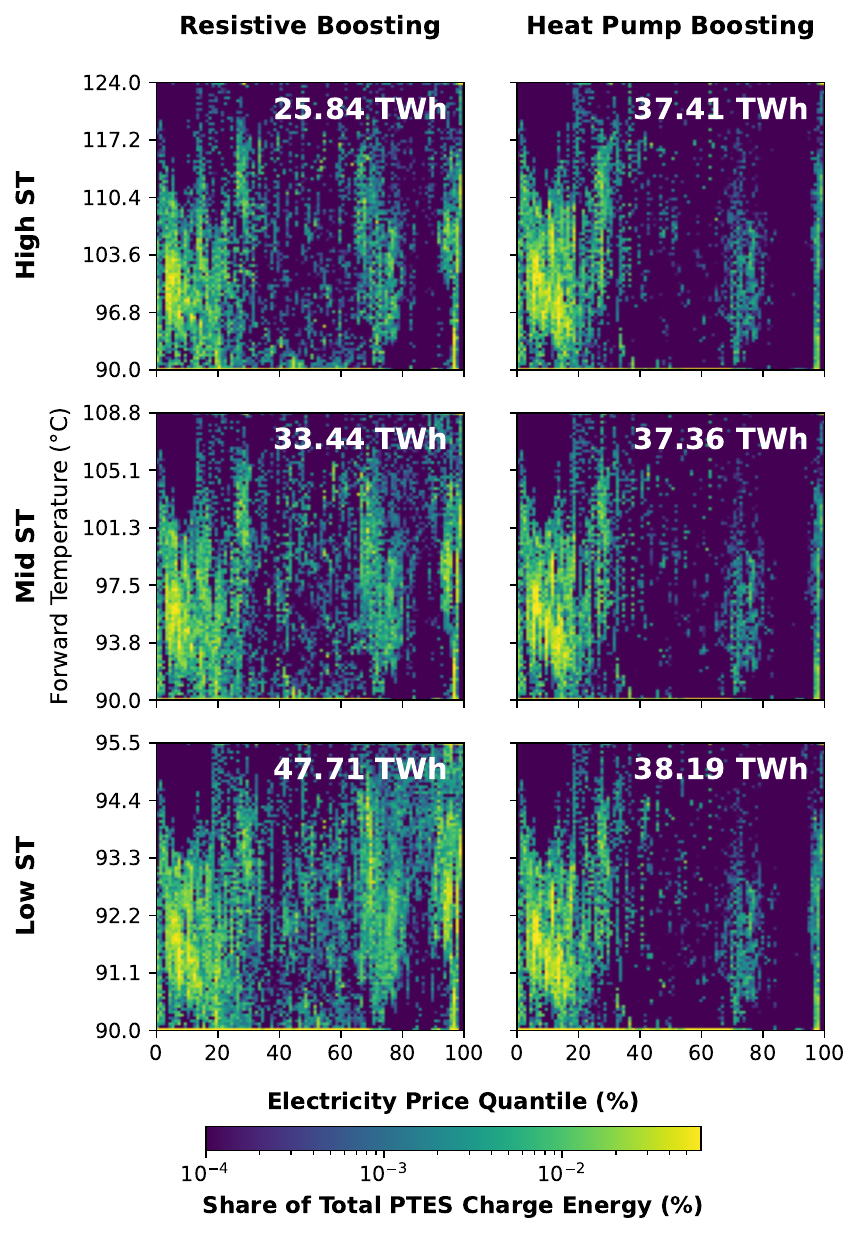}
    \end{minipage}
    \caption{Distribution of \ac{PTES} discharge (left) and charge (right) in all district heating systems over electricity price percentiles and district heating network forward temperatures. The distributions are shown for all three network temperature scenarios, both idealized scenarios, that do not require boosting and enlargen \ac{PTES} energy capacity respectively, and both feasible boosting configurations.}
    \label{fig:ptes_discharge_charge_by_price_and_T}
\end{figure*}

\begin{figure*}[!t]
    \centering
    \includegraphics[width=\textwidth]{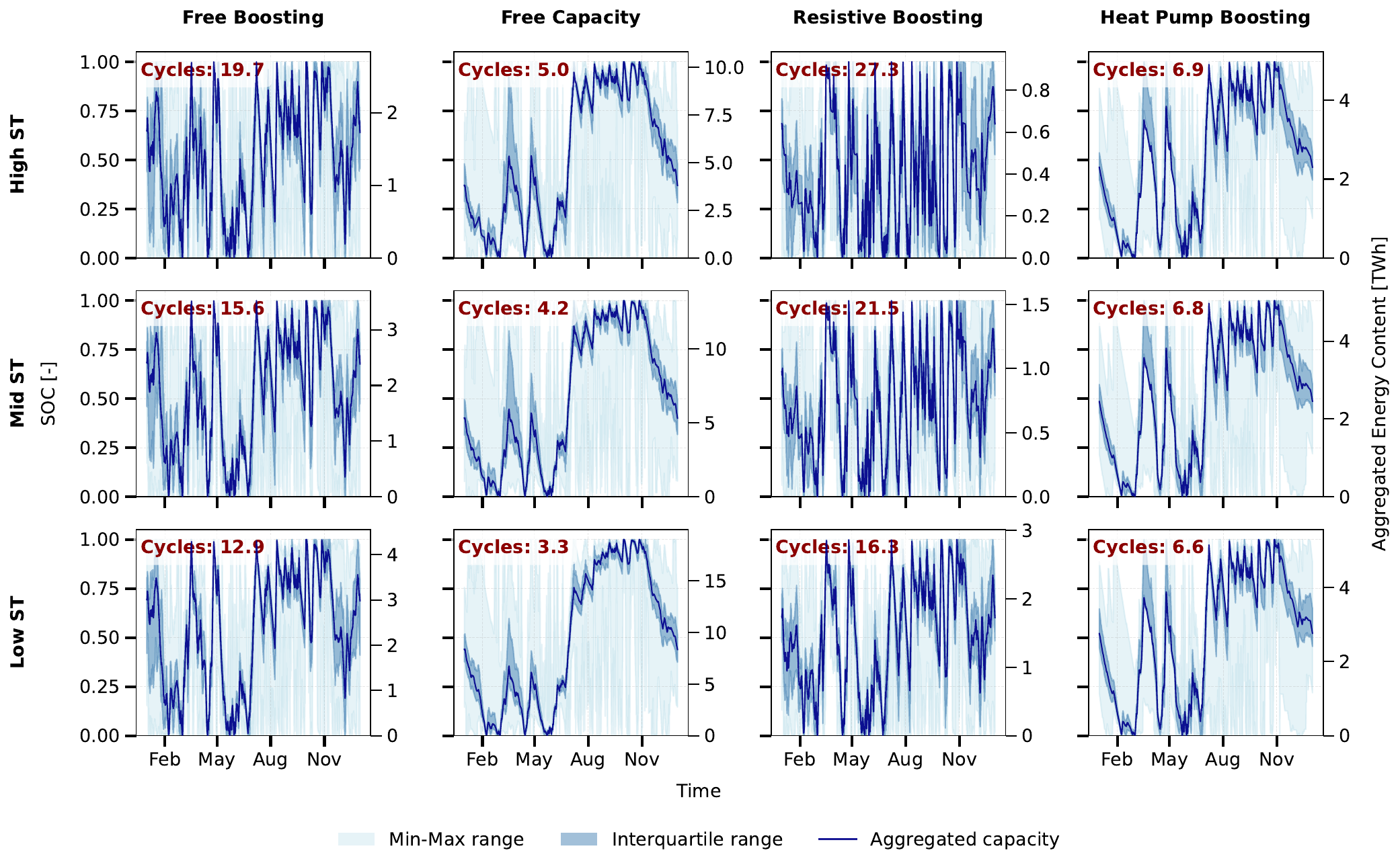}
    \caption{Yearly evolution of \ac{PTES} \ac{SOC} for different temperature levels and boosting configurations. The shaded areas represent the min-max range (light blue) and the interquartile range (medium blue) over all individual \ac{PTES} systems, while the dark blue line shows the aggregated \ac{SOC} across all systems. The corresponding absolute energy content of the aggregate \ac{PTES} is shown on the secondary y-axes. The number of cycles over the year shown in the top left corner of each subplot also refers to the aggregate \ac{PTES} capacity and discharge across all systems.}
    \label{fig:ptes_soc_comparison_distribution}
\end{figure*}

\end{document}